\documentclass[12pt]{article}
\usepackage{epsfig}
\usepackage{latexsym}
\usepackage{amsmath}
\usepackage{cite}
\numberwithin{equation}{section}

\title{
Graviton-Photon   Scattering\footnote{IPhT/t14/148, IHES/P/14/32}\\}
\author{\\ \hskip -.5cm  N.E.J. Bjerrum-Bohr$^a$, Barry R. Holstein$^b$, \\ Ludovic Plant\'e$^c$, Pierre Vanhove$^{c,d}$\\[1ex]
 $^a$ Niels Bohr International Academy and Discovery Center\\ The Niels Bohr Institute,
 Blegamsvej 17,
DK-2100 \\ Copenhagen \O, Denmark\smallskip\\[.5ex]
$^b$ Department of Physics-LGRT\\
University of Massachusetts\\
Amherst, MA  01003 USA\smallskip\\[.5ex]
$^c$ Institut de Physique Th\'eorique\\
CEA, IPhT, F-91191 Gif-sur-Yvette, France, and\\
CNRS, URA 2306, F-91191 Gif-sur-Yvette, France\smallskip\\[.5ex]
$^d$ Institut des Hautes \'Etudes Scientifiques,\\
F-91440 Bures-sur-Yvette, France}
\begin{document}
\begin{titlepage}
\maketitle
\vspace{-1cm}
\begin{abstract}

We use the feature that the gravitational Compton scattering amplitude
factorizes in terms of Abelian QED amplitudes to evaluate various gravitational Compton
processes. We examine both the QED and gravitational Compton
scattering from a massive spin-1 system by the use of helicity amplitude methods.
In the case of gravitational Compton scattering we show how the massless limit
can be used to evaluate the cross section for graviton-photon scattering
and discuss the difference between photon interactions and the zero mass
spin-1 limit.  We show that the forward scattering cross section for
graviton photoproduction has a very peculiar behavior,
differing from the standard Thomson and Rutherford cross sections for
a Coulomb-like potential.
\end{abstract}
\end{titlepage}

\section{Introduction}
The treatment of electromagnetic interactions in quantum mechanics is well known
and the discussion of electromagnetic effects via photon exchange is a staple of
the graduate curriculum. In particular photon exchange between charged
particles can be shown to give rise to the Coulomb potential as well
as to various higher order effects such as the spin-orbit and Darwin
interactions~\cite{bhb}. The fact that the photon carries spin-1
means that the electromagnetic current is a four-vector and
manipulations involving such vector quantities are familiar
to most physicists. In a similar fashion, graviton exchange
between a pair of masses can be shown to generate the
gravitational potential as well as various higher order
effects, but in this case the fact that the graviton is a
spin-2 particle means that gravitational ``currents"
are second rank tensors and the graviton propagator is a
tensor of rank four.  The resultant proliferation of indices is
one reason why this quantum mechanical discussion of
graviton exchange effects is not generally treated in introductory texts~\cite{msc}.

Recently, by the use of string-inspired methods, it has been demonstrated that
the gravitational interaction factorizes in such a way that a gravitational
amplitude can be written as the product of two more familiar vector
amplitudes~\cite{Kawai:1985xq,str,Bern:2002kj,Stieberger:2009hq,BjerrumBohr:2010hn}.
This factorization property, totally obscure at the level of the action, is a
fundamental properties of gravity and has deep consequences at the
loop amplitude level, since many gravitational amplitudes can be constructed
by an appropriate product of gauge theory integrand numerators~\cite{Bern:2010ue}.
This feature has triggered a good deal of new results in extended supergravity~\cite{Chiodaroli:2014xia,
Ochirov:2013xba,Bern:2009kd,Bern:2010yg,Bern:2012uf,Bern:2011rj,Carrasco:2012ca,
Bern:2012cd,Bern:2013yya,Bern:2013qca,Bern:2013uka,Bern:2014sna}, but quite
remarkably these techniques can be applied as well to pure
gravity~\cite{Johansson:2014zca,BjerrumBohr:2010hn}.

One remarkable property of amplitudes with emission of one or
two gravitons is their factorization in terms of \emph{Abelian}
QED amplitudes~\cite{gvp,BjerrumBohr:2010hn}.  This factorization property has the important
consequence that the low-energy limit of the gravitational Compton
amplitude for graviton photoproduction is directly connected
to the low-energy theorem for QED Compton amplitudes~\cite{BjerrumBohr:2010hn}.

In a previous paper~\cite{gvp} this property was used to
evaluate processes such as graviton photoproduction and
gravitational Compton scattering for both spin-0 and spin-${1\over 2}$
systems by simply evaluating the corresponding
electromagnetic amplitude for Compton scattering.  This simplification permits the
treatment of gravitational effects without long tedious computations,
since they are now no more difficult than corresponding electromagnetic
calculations.  The simplicity offered through factorization
has important consequences for the computations
of long-range corrections to interaction potentials containing loops of
intermediate photons- or gravitons~\cite{Holstein:2008sw,Holstein:2008sx,Holstein:2008sy,Bjerrum-Bohr:2013bxa}.
In this paper we extend such considerations to electromagnetic and gravitational interactions of spin-1
systems.  These calculations are useful not only as a generalization of our previous results but also, since the
photon carries spin one, such methods can be used to consider the case of photon-graviton scattering, although
there are subtleties in this case due to gauge invariance.

In all the cases under study, we show that the low-energy limit of the differential
cross section has a universal behavior independent of the spin of
the matter field on which photon or graviton is scattered.  We demonstrate that
this is a consequence of the well-known universal low-energy
behavior in quantum electrodynamics (QED) and the squaring relations
between gravitational and electromagnetic processes.

The forward differential cross section for the Compton scattering of
photons on a massive target has the well-known constant behavior of the
Thomson cross-section
\begin{equation}
\lim_{\theta_L\to 0}   {d\sigma^{\rm Comp}_{{\rm lab},S}\over d\Omega} ={\alpha^2\over2\,m^2}\,,
\end{equation}
while the small-angle limit of gravitational Compton scattering of gravitons on a massive
target has the expected behavior due to a $1/r$ long-range potential
of a Rutherford-like cross section
\begin{equation}
  \lim_{\theta_L\to0}
  {d\sigma^{\rm g-Comp}_{{\rm lab},S}\over d\Omega} = {16\,G^2\,m^2\over \theta_L^4}\,.
\end{equation}
We  explain in section~\ref{sec:forward} why this formula reproduces
the small-angle limit of the classical cross section for light bending
in a Schwarzschild background.

The forward limit of the graviton photoproduction cross section
has the rather unique behavior
\begin{equation} \label{e:fg0}
  \lim_{\theta_L\to0}   {d\sigma^{\rm photo}_{{\rm lab},S}\over d\Omega} = {4\,G\,\alpha\over \theta_L^2}\,.
\end{equation}
This limit is independent not only of the spin $S$ but as well of the
mass $m$ of the target. The small-angle dependence is typical of an
effective $1/r^2$ potential. We provide an explanation for this in
section~\ref{sec:forward}.

It may be very difficult to detect a single graviton~\cite{dyson} but photons
are easily detected so it is would be interesting to be able to use the
graviton photoproduction process to provide an indirect detection of a graviton.  The cross section in
Eq.~\eqref{e:fg0} is suppressed by a power of Newton's constant $G$ but,
being independent of the mass $m$ of the target, one can
discriminate this effect from that of Compton scattering.

In the next section then we quickly review the electromagnetic interaction and derive the spin-1 couplings.
In section~\ref{sec:compton}, we analyze the Compton scattering of a spin-1 particle.  The corresponding gravitational couplings are derived in section~\ref{sec:grav} and the graviton photoproduction and gravitational Compton scattering reactions are calculated via both direct and factorization methods. In section~\ref{sec:gravpho} we discuss photon-graviton scattering and the subtleties associated with gauge
invariance.  In section~\ref{sec:forward} we consider the forward small-angle
limit of the various scattering cross sections derived in the previous
section. We show that Compton, graviton photoproduction and the
gravitational Compton scattering have very different behavior.  We summarize our findings in a brief concluding section.

\section{Brief Review of Electromagnetism}\label{sec:review}

In this section we present a quick review of the electromagnetic and gravitational interactions and the results given in our previous work.  The electromagnetic interaction of a system may be found by using the minimal substitution $i\partial_\mu\rightarrow
iD_\mu=i\partial_\mu-eA_\mu$ in the free particle Lagrangian, where $A_\mu$ is the photon field.  In this way the Klein-Gordon Lagrangian
\begin{equation}
{\cal L}_0^{S=0}=\partial_\mu\phi^\dagger\,\partial^\mu\phi-m^2\,\phi^\dagger\phi\,,
\end{equation}
which describes a free charged spinless field, becomes
\begin{equation}
{\cal L}^{S=0}=(\partial_\mu
-ieA_\mu)\,\phi^\dagger(\partial^\mu+ieA^\mu)\,\phi+m^2\,\phi^\dagger\phi\,,
\end{equation}
after this substitution.  The corresponding interaction
Lagrangian can then be identified as
\begin{equation}
{\cal L}_{int}^{S=0}=ieA_\mu\phi^\dagger\overleftrightarrow{\nabla}^\mu\phi+e^2A^\mu
A_\mu\phi^\dagger\phi\,,
\end{equation}
where
\begin{equation}
C\overleftrightarrow{\nabla}D :=  C\vec{\nabla}D-(\vec{\nabla}D)C\,.
\end{equation}
Similarly, for spin-$\frac12$, the free Dirac Lagrangian
\begin{equation}
{\cal L}_0^{S={1\over 2}}=\bar{\psi}\,(i\not\!{\nabla}-m)\,\psi\,,
\end{equation}
becomes
\begin{equation}
{\cal L}^{S={1\over 2}}=\bar{\psi}\,(i\not\!{\nabla}-e\not\!\!{A}-m)\,\psi\,,
\end{equation}
and the interaction Lagrangian is found to be
\begin{equation}
{\cal L}_{int}^{S={1\over 2}}=-e\,\bar{\psi}\not\!\!{A}\,\psi\,.
\end{equation}
The resulting single-photon vertices are then
\begin{equation}
\big \langle p_f\,\big|V_{em}^{(1)\mu}\big|\,p_i\rangle_{S=0}=-i\,e\,(p_f+p_i)^\mu\,,
\end{equation}
for spin-0 and
\begin{equation}
\big\langle p_f\,\big|V_{em}^{(1)\mu}\big|\,p_i\big\rangle_{S={1\over 2}}=-i\,e\,\bar{u}(p_f)\gamma^\mu u(p_i)\,,
\end{equation}
for spin-$\frac12$, and in the case of spin 0 there exists also a two-photon ("seagull") vertex
\begin{equation}
\big\langle p_f\,\big|V_{em}^{(2){\mu\nu}}\big|\,p_i\big\rangle_{S=0}=2\,i\,e^2\eta^{\mu\nu}\,.
\end{equation}
The photon propagator in Feynman gauge is
\begin{equation}
D_f^{\alpha\beta}(q)={-i\eta^{\alpha\beta}\over q^2+i\epsilon}\,.
\end{equation}

The consequences of these Lagrangians were explored in ref. ~\cite{gvp} and in the present paper we extend
our considerations to the case of spin-1, for which the free Lagrangian
has the Proca form
\begin{equation}
{\cal L}_0^{S=1}=-{1\over 2}\,B^\dagger_{\mu\nu}B^{\mu\nu}+m^2\,B^\dagger_\mu B^\mu\,,
\end{equation}
where $B_\mu$ is a spin one field subject to the constraint $\partial^\mu B_\mu=0$ and $B_{\mu\nu}$ is the antisymmetric tensor
\begin{equation}
B^{\mu\nu}=\partial^\mu B^\nu-\partial^\nu B^\mu\,.
\end{equation}
The minimal substitution then leads to the interaction Lagrangian
\begin{equation}
{\cal L}_{int}^{S=1}=i\,e\,A^\mu B^{\nu\dagger}\left(\eta_{\nu\alpha}\overleftrightarrow{\nabla}_\mu-\eta_{\alpha\mu}\overleftrightarrow{\nabla}_\nu\right)B^\alpha
-e^2A^\mu A^\nu B^{\alpha\dagger}B^\beta(\eta_{\mu\nu}\eta_{\alpha\beta}-\eta_{\mu\alpha}\eta_{\nu\beta})\,,
\end{equation}
and the one, two photon vertices
\begin{eqnarray}
\big\langle p_f,\epsilon_B\,\big|V_{em}^{(1)\mu}\big|\,p_i,\epsilon_A\big\rangle_{S=1}&=&-i\,e\,\epsilon_{B\beta}^*\left((p_f+p_i)^\mu\eta^{\alpha\beta}-\eta^{\beta\mu}p^{f\alpha}
-\eta^{\alpha\mu}p^{i\beta}\right)\epsilon_{A\alpha}\,,\nonumber\\
\big\langle p_f,\epsilon_B\,\big|V_{em}^{(2)\mu\nu}\big|\,p_i,\epsilon_A\big\rangle_{S=1}&=&i\,e^2\,\epsilon_{B\beta}^*\left(2\eta^{\alpha\beta}\eta^{\mu\nu}
-\eta^{\alpha\mu}\eta^{\beta\nu}-\eta^{\alpha\nu}\eta^{\beta\mu}\right)\epsilon_{A\alpha}\,.\label{eq:kb}
\end{eqnarray}
However, Eq.~\eqref{eq:kb} is {\it not} the correct result for a fundamental spin-1 particle such as the charged $W$-boson.  Because the $W$ arises in a gauge theory, there exists an additional $W$-photon interaction, leading to an ``extra" contribution to the single photon vertex
\begin{equation}
\big\langle p_f,\epsilon_B\,\big|\delta V_{em}^{(1)\mu}\big|\,p_i,\epsilon_A\rangle_{S=1}=i\,e\,\epsilon_{B\beta}^*\left(\eta^{\alpha\mu}(p_i-p_f)^\beta-\eta^{\beta\mu}(p_i-p_f)^\alpha)\right)\epsilon_{A\alpha}\,.
\end{equation}
The meaning of this term can be seen by using the mass-shell Proca constraints $p_i\cdot\epsilon_A=p_f\cdot\epsilon_B=0$ to write the total on-shell single photon vertex as
\begin{eqnarray}
\big\langle p_f,\epsilon_B\,\big|(V_{em}+\delta V_{em})^\mu\big|\,p_i,\epsilon_A\big\rangle_{S=1}&=&-i\,e\,\epsilon_{B\beta}^*
\left((p_f+p_i)^\mu\eta_{\alpha\beta}-2\eta^{\alpha\mu}(p_i-p_f)^\beta\right.\nonumber\\
&+&\left.2\eta^{\beta\mu}(p_i-p_f)^\alpha\right)\epsilon_{A\alpha}\,,
\end{eqnarray}
wherein we observe that the coefficient of the term $-\eta^{\alpha\mu}(p_i-p_f)^\beta+\eta^{\beta\mu}(p_i-p_f)^\alpha$ has been modified from unity to two.  Since the rest frame spin operator can be identified via\footnote{Equivalently, one can use the relativistic identity
\begin{equation}
\epsilon^*_{B\mu}q\cdot\epsilon_A-\epsilon_{A\mu}q\cdot\epsilon_B^*={1\over 1-{q^2\over m^2}}\left({i\over m}\epsilon_{\mu\beta\gamma\delta}p_i^\beta q^\gamma S^\delta-{1\over 2m}(p_f+p_i)_\mu\epsilon_B^*\cdot q\epsilon_A\cdot q\right)
\end{equation}
where $S^\delta={i\over 2m}\epsilon^{\delta\sigma\tau\zeta}\epsilon^*_{B\sigma}\epsilon_{A\tau}(p_f+p_i)_\zeta$ is the spin four-vector.}
\begin{equation}
B^\dagger_i B_j-B^\dagger_j
B_i=-i\,\epsilon_{ijk}\big\langle f\,\big|S_k\big|\,i\big\rangle\label{eq:mm}\,,
\end{equation}
the corresponding piece of the nonrelativistic interaction Lagrangian becomes
\begin{equation}
{\cal L}_{\rm int}=-g\,{e\over 2m} \big\langle f\,\big|\vec{S}\big|\,i\big\rangle\cdot\vec{\nabla}\times\vec{A}\,,
\end{equation}
where $g$ is the gyromagnetic ratio  and we have included a factor
$2m$ which accounts for the normalization condition of the spin one
field.  Thus the ``extra" interaction required by a gauge theory
changes the $g$-factor from its Belinfante value of unity~\cite{bfn}
to its universal value of two, as originally proposed by Weinberg and
more recently buttressed by a number of arguments~\cite{wbg,large}.
Use of $g=2$ is required (as shown in~\cite{hbp}) in order to assure the validity of the
factorization result of gravitational amplitudes in terms of QED amplitudes, as used below.

\section{Compton Scattering}\label{sec:compton}

\begin{figure}
\begin{center}
\epsfig{file=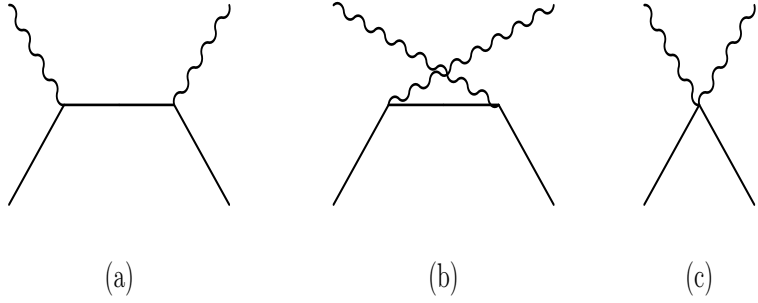,height=4cm,width=10cm} \caption{Diagrams
relevant to Compton scattering.}\label{fig:compton}
\end{center}
\end{figure}

The vertices given in the previous section can now be used to evaluate the Compton scattering amplitude for a spin-1 system having charge $e$ and mass $m$ by summing the contributions of the three diagrams shown in Figure~\ref{fig:compton}, yielding
\begin{eqnarray}
{\rm Amp}^{\rm Comp}_{S=1}
&=&e^2\Bigg[2\epsilon_A\cdot\epsilon_B^*\left[{\epsilon_i\cdot
p_i\epsilon_f^*\cdot p_f\over p_i\cdot k_i}-{\epsilon_i\cdot
p_f\epsilon_f^*\cdot p_i\over p_i\cdot
k_f}-\epsilon_i\cdot\epsilon_f^*\right]\nonumber\\
&-&g\left[\epsilon_A\cdot
[\epsilon_f^*,k_f]\cdot\epsilon_B^*\left({\epsilon_i\cdot p_i\over
p_i\cdot k_i} -{\epsilon_i\cdot p_f\over p_i\cdot k_f}\right)
- \epsilon_A\cdot[\epsilon_i,k_i]\cdot\epsilon_B^*\left({\epsilon_f\cdot
p_f\over p_i\cdot k_i}
-{\epsilon_f^*\cdot p_i\over p_i\cdot k_f}\right)\right]\nonumber\\
&-&g^2\left[{1\over 2p_i\cdot
k_i}\epsilon_A\cdot[\epsilon_i,k_i]\cdot[\epsilon_f^*,k_f]\cdot\epsilon_B^*
-{1\over 2p_i\cdot
k_f}\epsilon_A\cdot[\epsilon_f^*,k_f]\cdot[\epsilon_i,k_i]\epsilon_B^*\right]\nonumber\\
&-&{(g-2)^2\over m^2}\left[{1\over 2p_i\cdot
k_i}\epsilon_A\cdot[\epsilon_i,k_i]\cdot
p_i\epsilon_B^*\cdot[\epsilon_f^*,k_f]\cdot p_f\right.\nonumber\\
&-&\left.{1\over 2p_i\cdot
k_f}\epsilon_A\cdot[\epsilon_f^*,k_f]\cdot p_i
\epsilon_B^*\cdot[\epsilon_i,k_i]\cdot p_i\right]\Bigg]\,,\label{eq:gh}
\end{eqnarray}
with the momentum conservation condition $p_i+k_i=p_f+k_f$ and where we have defined
$$S\cdot[Q,R]\cdot T :=  S\cdot QT\cdot R-S\cdot RT\cdot Q.$$
We can verify the gauge invariance of the above form by noting that this amplitude
can be written in the equivalent form
\begin{eqnarray}
{\rm Amp}_{S=1}^{\rm Comp}&=&{e^2\over p_i\cdot k_ip_i\cdot k_f}\Bigg[2\epsilon_B^*\cdot\epsilon_A(p_i\cdot F_i\cdot F_f\cdot p_i)\nonumber\\
&+&\left.g\left[(\epsilon_B^*\cdot F_f\cdot\epsilon_A)(p_i\cdot F_i\cdot p_f)+(\epsilon_B^*\cdot F_i\cdot\epsilon_A)(p_i\cdot F_f\cdot p_f)\right]\right.\nonumber\\
&-&\left.{g^2\over 2}\left[p_i\cdot k_f(\epsilon_B^*\cdot F_f\cdot F_i\cdot \epsilon_A)-p_i\cdot k_i(\epsilon_B^*\cdot F_i\cdot F_f\cdot\epsilon_A)\right]\right.\nonumber\\
&-&{(g-2)^2\over 2m^2}\left[(\epsilon_B^*\cdot F_f\cdot p_f)(p_i\cdot F_i\cdot\epsilon_A)-(\epsilon_B^*\cdot F_i\cdot p_i)(p_i\cdot F_f\cdot \epsilon_A)\right]\Bigg]\,,\nonumber\\
\qquad
\end{eqnarray}
where $F_i^{\mu\nu}=\epsilon_i^\mu k_i^\nu-\epsilon_i^\nu k_i^\mu$ and $F_f^{\mu\nu}=\epsilon_f^{*\mu} k_f^\nu-\epsilon_f^{*\nu} k_f^\mu$.  Since $F_{i,f}$ are obviously invariant under the substitutions $\epsilon_{i,f}\rightarrow \epsilon_{i,f}+\lambda k_{i,f},\,\,i=1,2$, it is clear that Eq.~\eqref{eq:gh} satisfies the gauge invariance strictures
\begin{equation}
\epsilon_f^{*\mu}k_i^\nu{\rm Amp}^{\rm Comp}_{\mu\nu,S=1}=k_f^\mu\epsilon_i^\nu{\rm Amp}^{\rm Comp}_{\mu\nu,S=1}=0\,.
\end{equation}
Henceforth in this manuscript we shall assume the $g$-factor of the
spin-1 system to have its ``natural'' value $g=2$, since it is in this
case that the high-energy properties of the scattering are well
controlled and the factorization methods of gravity amplitudes are valid~\cite{wbg,large}.

In order to make the transition to gravity in the next section, it is useful to utilize the helicity formalism~\cite{hct}, whereby we evaluate the matrix elements of the Compton amplitude between initial and final spin-1 and photon states having definite helicity, where helicity is defined as the projection of the particle spin along the momentum direction. We shall work initially in the center of mass frame.  For a photon incident with four-momentum $k_{i\mu}=p_{\rm CM}(1,\hat{z})$ we choose the polarization vectors,
\begin{equation}
\epsilon_i^{\lambda_i}=-{\lambda_i\over
\sqrt{2}}\big(\hat{x}+i\lambda_i \hat{y}\big),\qquad \lambda_i=\pm\,,
\end{equation}
while for an outgoing photon with $k_{f\mu}=p_{\rm CM}(1,\cos\theta_{\rm CM}\hat{z}+\sin\theta_{\rm CM}\hat{x})$ we use polarizations
\begin{equation}
\epsilon_f^{\lambda_f}=-{\lambda_f\over
\sqrt{2}}\Big(\cos\theta_{\rm CM}\hat{x}+
i\lambda_f\hat{y}-\sin\theta_{\rm CM}\hat{z}\Big),\quad \lambda_f=\pm\,.
\end{equation}
We can define corresponding helicity states for the spin-1 system.  In this case the initial and final four-momenta are
$p_i=(E_{\rm CM},-p_{\rm CM}\hat{z})$ and $p_f=(E_{\rm CM},-p_{\rm CM}(\cos\theta_{\rm CM}\hat{z}+\sin\theta_{\rm CM}\hat{x}))$ and there are transverse polarization four-vectors
\begin{eqnarray}
\epsilon_{A\mu}^{\pm}&=&\Big(0,\mp{-\hat{x}\pm i\hat{y}\over \sqrt2}\Big)\,,\nonumber\\
\epsilon_{B\mu}^{\pm}&=&\Big(0,\mp{-\cos\theta_{\rm CM}\hat{x}\pm i\hat{y}+\sin\theta_{\rm CM}\hat{z}\over \sqrt 2}\Big)\,,
\end{eqnarray}
while the longitudinal mode has polarization four-vectors
\begin{eqnarray}
\epsilon_{A\mu}^0&=&{1\over m}\big(p_{\rm CM},-E_{\rm CM}\hat{z}\big)\,,\nonumber\\
\epsilon_{B\mu}^0&=&{1\over m}\big(p_{\rm CM},-E_{\rm CM}(\cos\theta_{\rm CM}\hat{z}+\sin\theta_{\rm CM}\hat{x})\big)\,,
\end{eqnarray}
In terms of the usual invariant kinematic variables
$$s=\big(p_i+k_i\big)^2,\quad t=\big(k_i-k_f\big)^2,\quad u=\big(p_i-k_f\big)^2\,,$$
we identify
\begin{eqnarray}
p_{\rm CM}&=&{s-m^2\over 2\sqrt{s}}\,,\nonumber\\
E_{\rm CM}&=&{s+m^2\over 2\sqrt{s}}\,,\nonumber\\
\cos{1\over
2}\theta_{\rm CM}&=&{\big((s-m^2)^2+s t\big)^{1\over 2}\over s-m^2}={\big(m^4-s u\big)^{1\over 2}\over s-m^2}\,,\nonumber\\
\sin{1\over 2}\theta_{\rm CM}&=&{\big(-s t\big)^{1\over 2}\over s-m^2}\,.\label{eq:bv}
\end{eqnarray}
The invariant cross section for unpolarized Compton scattering is then given by
\begin{equation}
{d\sigma_{S=1}^{\rm Comp}\over dt}={1\over 16\pi(s-m^2)^2}\ {1\over
3}\sum_{a,b=-,0,+}{1\over 2}\sum_{c,d=-,+}\Big|B^1(ab;cd)\Big|^2\,.\label{eq:nh}\displaystyle
\end{equation}
where
\begin{equation}
B^1(ab;cd)=\big \langle p_f,b;k_f,d\,\big | {\rm Amp}_{S=1}^{\rm Comp}\big | \,p_i,a;k_i,c\,\big\rangle\,,
\end{equation}
is the Compton amplitude for scattering of a photon with four-momentum $k_i$, helicity a from a spin-1 target having four-momentum $p_i$, helicity $c$ to a photon with four-momentum $k_f$, helicity $d$ and target with four-momentum $p_f$, helicity $b$.  The helicity amplitudes can be calculated straightforwardly.  There exist $3^2\times2^2=36$ such amplitudes but, since helicity reverses under spatial inversion, parity invariance of the electromagnetic interaction requires that\footnote{Note that we require only that the magnitudes of the helicity amplitudes related by parity and/or time reversal be the same.  There could exist unobservable phases.}
$$\big|B^1(ab;cd)\big|=\big|B^1(-a-b;-c-d)\big|\,.$$
Also, since helicity is unchanged under time reversal, but initial and final states are interchanged, time reversal invariance of the electromagnetic interaction requires that
$$\big|B^1(ab;cd)\big|=\big|B^1(ba;dc)\big|\,.$$
Consequently there exist only twelve {\it independent} helicity amplitudes.  Using Eq.~\eqref{eq:gh} we can calculate the various helicity amplitudes in the center of mass frame and then write these results in terms of invariants using Eq.~\eqref{eq:bv}, yielding
\begin{eqnarray}
\big |B^1(++;++)\big |&=&\big |B^1(--;--)\big |=2e^2{\big((s-m^2)^2+m^2t\big)^2\over (s-m^2)^3(u-m^2)}\,,\nonumber\\
\big |B^1(++;--)\big |&=&\big |B^1(--;++)\big |=2e^2{(m^4-su)^2\over (s-m^2)^3 (u-m^2)}\,,\nonumber\\
\big |B^1(+-;+-)\big |&=&\big |B^1(-+;-+)\big |=2e^2{m^4t^2\over (s-m^2)^3(u-m^2)}\,,\nonumber\\
\big |B^1(+-;-+)\big |&=&\big |B^1(-+;+-)\big |=2e^2{s^2t^2\over(s-m^2)^3(u-m^2)}\,,\nonumber\\
\big |B^1(++;+-)\big |&=&\big |B^1(--;-+)\big |=\big|B^1(++;-+)\big|=\big|B^1(--;+-)\big|\,,\nonumber\\
&=&2e^2{m^2t(m^4-su)\over(s-m^2)^3(u-m^2)}\,,\nonumber\\
\big|B^1(+-;++)\big|&=&\big|B^1(-+;--)\big|=\big|B^1(-+;++)\big|=\big|B^1(+-;--)\big|\,,\nonumber\\
&=&2e^2{m^2t(m^4-su)\over (s-m^2)^3(u-m^2)}\,.
\end{eqnarray}
and
\begin{eqnarray}
\big|B^1(0+;++)\big|&=&\big|B^1(0-;--)\big|=\big|B^1(+0;++)\big|=\big|B^1(-0;--)\big|\,,\nonumber\\
&=&2e^2{\sqrt{2}m\big(t m^2+(s-m^2)^2\big)\sqrt{-t(m^4-su)}\over (s-m^2)^3(u-m^2)}\,,\nonumber\\
\big|B^1(0+;+-)\big|&=&\big|B^1(0-;-+)\big|=\big|B^1(+0;-+)\big|=\big|B^1(-0;+-)\big|\,,\nonumber\\
&=&2e^2{\sqrt{2}mst\sqrt{-t(m^4-su)}\over (s-m^2)^3(u-m^2)}\,,\nonumber\\
\big|B^1(0+;-+)\big|&=&\big|B^1(0-;+-)\big|=\big|B^1(+0;+-)\big|=\big|B^1(-0;-+)\big|\,,\nonumber\\
&=&2e^2{\sqrt{2}m^3t\sqrt{-t(m^4-su)}\over (s-m^2)^3(u-m^2)}\,,\nonumber\\
\big|B^1(0+;--)\big|&=&\big|B^1(0-;++)\big|=\big|B^1(+0;--)\big|=\big|B^1(-0;++)\big|\,,\nonumber\\
&=&2e^2{\sqrt{2}m\big(-t(m^4-su)\big)^{3\over 2}\over (s-m^2)^3t(u-m^2)}\,,\nonumber\\
\big|B^1(00;++)\big|&=&\big|B^1(00;--)\big|=2e^2{\big(2tm^2+(s-m^2)^2\big)(m^4-su)\over (s-m^2)^3(u-m^2)}\,,\nonumber\\
\big|B^1(00;+-)\big|&=&\big|B^1(00;-+)\big|=2e^2{\big(m^2t((s-m^2)^2+2st\big)\over (s-m^2)^3(u-m^2)}\,.
\end{eqnarray}
Substitution into Eq.~\eqref{eq:nh} then yields the invariant cross section for unpolarized Compton scattering from a spin-1 target
\begin{equation}
{d\sigma^{\rm Comp}_{S=1}\over dt}={e^4\over 12\pi(s-m^2)^4(u-m^2)^2}\Big[(m^4-su+t^2)\big(3(m^4-su)+t^2\big)+t^2(t-m^2)(t-3m^2)\Big]\,,
\end{equation}
which can be compared with the corresponding results for unpolarized Compton scattering from spin-0 and spin-$\frac12$ targets found in ref.~\cite{gvp}---
\begin{eqnarray}
{d\sigma_{S=0}^{\rm Comp}\over dt}&=&{e^4\over 4\pi(s-m^2)^4(u-m^2)^2}\left[(m^4-su)^2+m^4t^2\right]\,,\nonumber\\
{d\sigma_{S={1\over 2}}^{\rm Comp}\over dt}&=&{e^4\Big[(m^4-su)\big(2(m^4-su)+t^2\big)+m^2t^2(2m^2-t)\Big]\over 8\pi{(s-m^2)^4(u-m^2)^2}}\,.
\end{eqnarray}
Usually such results are written in the {\it laboratory} frame, wherein the target is at rest, by use of the relations
\begin{eqnarray}
s-m^2&=&2m\omega_i,\quad u-m^2=-2m\omega_f\,,\nonumber\\[1ex]
m^4-su&=&4m^2\omega_i\omega_f\cos^2{\theta_L\over 2},\quad
m^2t=-4m^2\omega_i\omega_f\sin^2{\theta_L\over 2}\,,
\end{eqnarray}
and
\begin{equation}
{dt\over d\Omega}={d\over 2\pi
d\cos\theta_L}\left(-{2\omega_i^2(1-\cos\theta_L)\over
1+{\omega_i\over m}(1-\cos\theta_L)}\right)={\omega_f^2\over \pi}\,.
\end{equation}
Introducing the fine structure constant $\alpha=e^2/4\pi$, we find then
\begin{eqnarray}
{d\sigma_{{\rm lab},S=1}^{\rm Comp}\over d\Omega}&=&{\alpha^2\over m^2}{\omega_f^4\over \omega_i^4}\left[\bigg(\cos^4{\theta_L\over 2}+\sin^4{\theta_L\over 2}\bigg)\bigg(1+2{\omega_i\over m}\sin^2{\theta_L\over 2}\bigg)^2\right.\nonumber\\
&+&\left.{16\omega_i^2\over 3m^2}\sin^4{\theta_L\over 2}\bigg(1+2{\omega_i\over m}\sin^2{\theta_L\over 2}\bigg)+{32\omega_i^4\over 3m^4}\sin^8{\theta_L\over 2}\right]\,,\nonumber\\
{d\sigma^{\rm Comp}_{{\rm lab},S={1\over 2}}\over d\Omega}&=&{\alpha^2\over m^2}{\omega_f^3\over \omega_i^3}\left[\bigg(\cos^4{\theta_L\over 2}+\sin^4{\theta_L\over 2}\bigg)\bigg(1+2{\omega_i\over m}\sin^2{\theta_L\over 2}\bigg)+2{\omega_i^2\over m^2}\sin^4{\theta_L\over 2}\right]\,,\nonumber\\
{d\sigma^{\rm Comp}_{{\rm lab},S=0}\over d\Omega}&=&{\alpha^2\over m^2}{\omega_f^2\over\omega_i^2}\left[\cos^4{\theta_L\over 2}+\sin^4{\theta_L\over 2}\right]\,.
\end{eqnarray}
We observe that the nonrelativistic laboratory cross section has an identical form for {\it any} spin
\begin{equation}\label{e:NRphoto}
\left.{d\sigma_{{\rm lab},S}^{\rm Comp}\over d\Omega}\right|^{NR}={\alpha^2\over 2m^2}\left[\bigg(\cos^4{\theta_L\over 2}+\sin^4{\theta_L\over 2}\Big)\Big(1+{\cal O}\Big({\omega_i\over m}\Big)\bigg)\right]\,,
\end{equation}
which follows from the universal form of the Compton amplitude for scattering from a spin-$S$ target in the low-energy ($\omega\ll m$) limit, which in turn arises from the universal form of the Compton amplitude for scattering from a spin-$S$ target in the low-energy limit---
\begin{equation}
\big\langle S,M_f;\epsilon_f\,\big|{\rm Amp}^{\rm Comp}_S\big|\,S,M_i;\epsilon_i\big\rangle_{\omega\ll m}=2e^2\,\epsilon_f^*\cdot\epsilon_i\,\delta_{M_i,M_f}+\ldots\,,
\end{equation}
and obtains in an effective field theory approach to Compton
scattering~\cite{wsb}.\footnote{ That the seagull contribution
dominates the nonrelativistic cross section is clear from the feature that
\begin{equation}
{\rm Amp}_{\rm Born}\sim 2e^2{\epsilon_f^*\cdot p\epsilon_i\cdot p\over p\cdot k}\sim{\omega\over m}\times{\rm Amp}_{\rm seagull}=2e^2\epsilon_f^*\cdot\epsilon_i.
\end{equation}}

\section{Gravitational Interactions}\label{sec:grav}

In the previous section we discussed the treatment the familiar electromagnetic interaction, using Compton scattering on a spin-1 target as an example.  In this section we show how the gravitational interaction can be evaluated via methods parallel to those used in the electromagnetic case.  An important difference is that while in the electromagnetic case we have the simple interaction Lagrangian
\begin{equation}
{\cal L}_{int}=-eA_\mu J^\mu\,,
\end{equation}
where $J^\mu$ is the electromagnetic current matrix element, for gravity we have
\begin{equation}
{\cal L}_{int}=-{\kappa\over 2}h^{\mu\nu}T^{\mu\nu}\,.
\end{equation}
Here the field tensor $h_{\mu\nu}$ is defined in terms of the metric via
\begin{equation}
g_{\mu\nu}=\eta_{\mu\nu}+\kappa h_{\mu\nu}\,,
\end{equation}
where $\kappa$ is given in terms of Newton's constant by
$\kappa^2=32\pi G$. The Einstein-Hilbert action is
\begin{equation}
  \label{e:EH}
  \mathcal S_{\rm Einstein-Hilbert}=\int d^4x \,\sqrt{-g} \, {2\over
    \kappa^2}\, \mathcal R\,,
\end{equation}
where
\begin{equation}
\sqrt{-g}=\sqrt{-{\rm det}g}=\exp{1\over 2}{\rm tr}{\rm log} g=1+{1\over 2}\eta^{\mu\nu}h_{\mu\nu}+\ldots\,,
\end{equation}
is the square root of the determinant of the metric and $\mathcal R:=
R^\lambda{}_{\mu\lambda\nu} g^{\mu\nu}$ is the Ricci scalar curvature obtained by contracting the Riemann tensor
$R^\mu{}_{\nu\rho\sigma}$ with the metric tensor.  The energy-momentum tensor is defined in
terms of the matter Lagrangian via
\begin{equation}
T_{\mu\nu}={2\over \sqrt{-g}}{\delta {\sqrt{-g}\cal L}_{\rm mat}\over
\delta g^{\mu\nu}}\,.\label{eq:pk}
\end{equation}
The prescription Eq.~\eqref{eq:pk} yields the forms
\begin{equation}
T_{\mu\nu}^{S=0}=\partial_\mu\phi^\dagger\partial_\nu\phi+\partial_\nu\phi^\dagger
\partial_\mu\phi-g_{\mu\nu}(\partial_\lambda\phi^\dagger\partial^\lambda\phi-m^2\phi^\dagger\phi)\,,
\end{equation}
for a scalar field and
\begin{equation}
T_{\mu\nu}^{S=\frac12}=\bar{\psi}[{1\over 4}\gamma_\mu
i\overleftrightarrow{\nabla}_\nu+{1\over 4}\gamma_\nu
i\overleftrightarrow{\nabla}_\mu-g_{\mu\nu}({i\over
2}\ \slash\!\!\!\!\!{\overleftrightarrow{\nabla}}-m)]\psi\,,
\end{equation}
for spin-$\frac12$, where we have defined
\begin{equation}
\bar{\psi}i\overleftrightarrow{\nabla}_\mu\psi :=
\bar{\psi}i\nabla_\mu\psi-(i\nabla_\mu\bar{\psi})\psi\,.
\end{equation}
\begin{figure}
\begin{center}
\epsfig{file=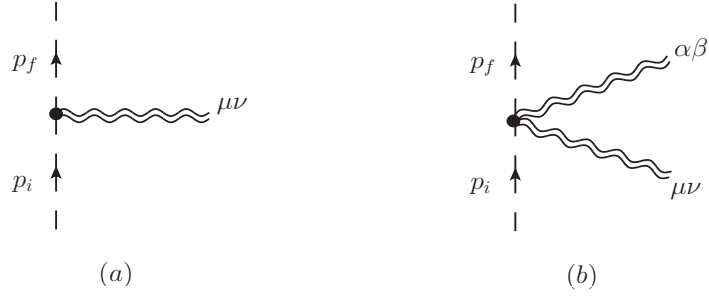,height=4cm,width=10cm} \caption{ (a) The one-graviton and (b) two-graviton emission vertices from either a
  scalar, spinor or vector particle. }\label{fig:vertices}
\end{center}
\end{figure}
\begin{figure}
\begin{center}
  \epsfig{file=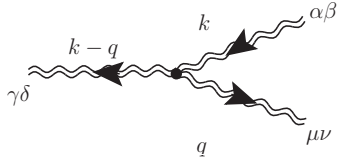,width=5cm} \caption{ The three
  graviton vertex }\label{fig:3grav}
\end{center}
\end{figure}

The one graviton emission vertices of figure~\ref{fig:vertices}(a) can now be
identified as
\begin{equation}
\big\langle p_f\,\big|V_{grav}^{(1)\mu\nu}\big|\,p_i\big\rangle_{S=0}=-i\,{\kappa\over 2}\,\left[p_f^\mu p_i^\nu +p_f^\nu p_i^\mu-\eta^{\mu\nu}\big(p_f\cdot
p_i-m^2\big)\right]\,,
\end{equation}
for spin-0,
\begin{equation}
\big\langle p_f\,\big|V_{grav}^{(1)\mu\nu}\big|\,p_i\big\rangle_{S={1\over 2 }}=-i\,{\kappa\over 2}\,\bar{u}(p_f)\bigg[{1\over 4
}\gamma^\mu\big(p_f+p_i\big)^\nu+{1\over 4}\gamma^\nu\big(p_f+p_i\big)^\mu\bigg]u(p_i)\,,
\end{equation}
for spin-$\frac12$, and
\begin{eqnarray}
\big\langle p_f,\epsilon_B\,\big|V_{grav}^{(1)\mu\nu}\big|\,p_i,\epsilon_A\big\rangle_{S=1}&=&-i\,{\kappa\over 2}\,\bigg[\epsilon_B^*\cdot\epsilon_A\big(p_i^\mu p_f^\nu+p_i^\nu p_f^\mu\big)-\epsilon_B^*\cdot p_i\,\big(p_f^\mu\epsilon_A^\nu+\epsilon_A^\mu p_f^\nu\big)\nonumber\\
& -&\epsilon_A\cdot p_f\big(p_i^\nu\epsilon_B^{*\mu}+p_i^\mu\epsilon_B^{*\nu}\big)+\big(p_f\cdot p_i-m^2\big)\big(\epsilon_A^\mu\epsilon_B^{*\nu}+\epsilon_A^\nu\epsilon_B^{*\mu}\big)\nonumber\\
& -&\eta^{\mu\nu}\left[\big(p_i\cdot p_f-m^2\big)\epsilon_B^*\cdot\epsilon_A-\epsilon_B^*\cdot p_i\,\epsilon_A\cdot p_f\right]\bigg]\,,\label{eq:dt}
\end{eqnarray}
for spin-1.  There also exist two-graviton (seagull) vertices shown in figure~\ref{fig:vertices}(b), which
can be found by expanding the stress-energy tensor to second order in $h_{\mu\nu}$.  In the case of spin-0
\begin{eqnarray}
\big\langle p_f\,\big|V_{grav}^{(2)\mu\nu,\alpha\beta}\big|\,p_i\big\rangle_{S=0}&=&i\kappa^2\left[{I^{\mu\nu,}}_{\rho\xi}{I^\xi}_{\zeta,\alpha\beta}(p_f^\zeta p_i^\rho+p_f^\rho p_i^\zeta)
-{1\over 2}\left(\eta^{\mu\nu}I^{\rho\zeta,\alpha\beta}+\eta^{\alpha\beta}I^{\rho\zeta,\mu\nu}p_f^\rho p_i^\zeta\right)\right.\nonumber\\
& -&\left.{1\over 2}\left(I^{\mu\nu;\alpha\beta}-{1\over 2}\eta^{\mu\nu}\eta^{\alpha\beta}\right)\big(p_f\cdot p_i-m^2\big)\right]\,,
\end{eqnarray}
where
\begin{equation}I_{\alpha\beta,\gamma\delta}={1\over
    2}\big(\eta_{\alpha\gamma}\eta_{\beta\delta}+\eta_{\alpha\delta}\eta_{\beta\gamma}\big).\end{equation}
For spin-$\frac12$
\begin{eqnarray}
\big\langle p_f\,\big|V_{grav}^{(2)\mu\nu,\alpha\beta}\big|\,p_i\big\rangle_{S=\frac12}&=&i\kappa^2\,\bar
u(p_f)\bigg[
{1\over
16}\Big[\eta^{\mu\nu}\big(\gamma^\alpha(p_f+p_i)^\beta+\gamma^\beta(p_f+p_i)^\alpha\big)
\nonumber\\
&+&\eta^{\alpha\beta}\big(\gamma^\mu
(p_f+p_i)^\nu+\gamma^\nu(p_f+p_i)^\mu\big)\Big]\nonumber\\
&+&{3\over
16}(p_f+p_i)_{\epsilon}\gamma_{\xi}\big(I^{\xi\phi,\mu\nu}{I_{\phi}}^{\epsilon,\alpha\beta}
+I^{\xi\phi,\alpha\beta}{I_{\phi}}^{\epsilon,\mu\nu}\big)\nonumber\\
&+&{i\over 16}\epsilon^{\rho\sigma\eta\lambda}\gamma_\lambda
\gamma_5\big({I^{\mu\nu,\eta}}_\zeta
I^{\alpha\beta,\sigma\zeta}p_{f\rho}-{I^{\alpha\beta,\eta}}_\zeta
I^{\mu\nu,\sigma\zeta}p_{i\rho}\big)\bigg]\, u(p_i)\,,\nonumber\\
\quad
\end{eqnarray}
while for spin-1
\begin{eqnarray}
\big\langle p_f, \epsilon_B;k_f\,\big|V_{grav}^{(2)\,\mu\nu,\rho\sigma}\big|\,p_i,\epsilon_A;k_i\big\rangle_{S=1}\!\!\!\!&=&-i\,{\kappa^2\over
4}\,\bigg[\nonumber\\ \nonumber&+&\big[p_{i\beta}p_{f\alpha}
-\left.
          \eta_{\alpha\beta}(p_i\cdot p_f - m^2)\big]
      \big(\eta_{\mu\rho}\eta_{\nu\sigma}+
          \eta_{\mu\sigma}\eta_{\nu\rho} -
          \eta_{\mu\nu}\eta_{\rho\sigma}\big)\right.\nonumber\\
          &+&\left.
    \eta_{\mu\rho}\big[\eta_{\alpha\beta}\big(p_{i\nu}p_{f\sigma} +
                p_{i\sigma}p_{f\nu}\big) -
          \eta_{\alpha\nu}p_{i\beta}p_{f\sigma}-
          \eta_{\beta\nu}p_{i\sigma}p_{f\alpha}\right.\nonumber\\
          &-&\left.
          \eta_{\beta\sigma}p_{i\nu}p_{f\alpha} -
          \eta_{\alpha\sigma}p_{i\beta}
            p_{f\nu} + (p_i\cdot p_f -
                m^2\big)\big(\eta_{\alpha\nu}\eta_{\beta\sigma} +
                \eta_{\alpha\sigma}\eta_{\beta\nu}\big)\big]\right.\nonumber\\
                &+&\left.
    \eta_{\mu\sigma}\big[\eta_{\alpha\beta}\big(p_{i\nu}p_{f\rho} +
                p_{i\rho}p_{f\nu}\big) -
          \eta_{\alpha\nu}p_{i\beta}p_{f\rho} -
          \eta_{\beta\nu}p_{i\rho}p_{f\alpha}\right.\nonumber\\
          &-&\left.
          \eta_{\beta\rho}p_{i\nu}p_{f\alpha}-
          \eta_{\alpha\rho}p_{i\beta}
            p_{f\nu} + (p_i\cdot p_f -
                m^2\big)\eta_{\alpha\nu}\eta_{\beta\rho} +
                \eta_{\alpha\rho}\eta_{\beta\nu}\big)\big]\right.\nonumber\\
                &+&\left.
    \eta_{\nu\rho}\big[\eta_{\alpha\beta}\big(p_{i\mu}p_{f\sigma} +
                p_{i\sigma}p_{f\mu}\big)
                -\eta_{\alpha\mu}p_{i\beta}p_{f\sigma} -
          \eta_{\beta\mu}p_{i\sigma}p_{f\alpha}\right.\nonumber\\
          &-&\left.\eta_{\beta\sigma}p_{i\mu}p_{f\alpha}
          -\eta_{\alpha\sigma}p_{i\beta}
            p_{f\mu} + (p_i\cdot p_f -
                m^2\big)\big(\eta_{\alpha\mu}\eta_{\beta\sigma} +
                \eta_{\alpha\sigma}\eta_{\beta\mu}\big)\big]\right.\nonumber\\
                &+&\left.
    \eta_{\nu\sigma}\big[\eta_{\alpha\beta}\big(p_{i\mu}p_{f\rho} +
                p_{i\rho}p_{f\mu}\big) -
          \eta_{\alpha\mu}p_{i\beta}p_{f\rho} -
          \eta_{\beta\mu}p_{i\rho}p_{f\alpha}\right.\nonumber\\
          &-&\left.\eta_{\beta\rho}p_{i\mu}p_{f\alpha}-\eta_{\alpha\rho}p_{i\beta}
            p_{f\mu} + (p_i\cdot p_f -
                m^2\big)\big(\eta_{\alpha\mu}\eta_{\beta\rho} +
                \eta_{\alpha\rho}\eta_{\beta\mu}\big)\big]\right.\nonumber\\
                &-&\left.
    \eta_{\mu\nu}\big[\eta_{\alpha\beta}\big(p_{i\rho}p_{f\sigma} +
                p_{i\sigma}p_{f\rho}\big) -
          \eta_{\alpha\rho}p_{i\beta}p_{f\sigma} -
          \eta_{\beta\rho}p_{i\sigma}p_{f\alpha}\right.\nonumber\\
          &-&\left.\eta_{\beta\sigma}p_{i\rho}p_{f\alpha}-
          \eta_{\alpha\sigma}p_{i\beta}p_{f\rho} + \big(p_i\cdot p_f -
                m^2\big)\big(\eta_{\alpha\rho}\eta_{\beta\sigma} +
                \eta_{\beta\rho}\eta_{\alpha\sigma}\big)\big]\right.\nonumber\\
                &-&\left.
    \eta_{\rho\sigma}\big[\eta_{\alpha\beta}\big(p_{i\mu}p_{f\nu} +
                p_{i\nu}p_{f\mu}\big) -
          \eta_{\alpha\mu}p_{i\beta}p_{f\nu} -
          \eta_{\beta\mu}p_{i\nu}p_{f\alpha}\right.\nonumber\\
          &-&\left.
          \eta_{\beta\nu}p_{i\mu}p_{f\alpha} -
          \eta_{\alpha\nu}p_{i\beta}
            p_{f\mu} + (p_i\cdot p_f -
                m^2\big)\big(\eta_{\alpha\mu}\eta_{\beta\nu} +
                \eta_{\beta\mu}\eta_{\alpha\nu}\big)\big]\right.\nonumber\\
                 &+&\left.
    \big(\eta_{\alpha\rho}p_{i\mu} -
          \eta_{\alpha\mu}p_{i\rho}\big)\big(\eta_{\beta\sigma}
            p_{f\nu} - \eta_{\beta\mu}p_{f\sigma}\big)\right.\nonumber\\
            &+&\left.
    \big(\eta_{\alpha\sigma}p_{i\nu} -
          \eta_{\alpha\nu}p_{i\sigma}\big)\eta_{\beta\rho}
            p_{f\mu} - \eta_{\beta\mu}p_{f\rho}\big)\right.\nonumber\\
            &+&\left.
    \big(\eta_{\alpha\sigma}p_{i\mu} -
          \eta_{\alpha\mu}p_{i\sigma}\big)\big(\eta_{\beta\rho}
          p_{f\nu} - \eta_{\beta\nu}p_{f\rho}\big)\right.\nonumber\\
          &+&\left.
    \big(\eta_{\alpha\rho}p_{i\nu} -
          \eta_{\alpha\nu}p_{i\rho}\big)\big(\eta_{\beta\sigma}
            p_{f\mu} - \eta_{\beta\mu}p_{f\sigma}\big)\right\}\epsilon_A^\alpha(\epsilon_B^\beta)^*\,.
\end{eqnarray}

Finally, we require the triple graviton vertex of figure~\ref{fig:3grav}
\begin{eqnarray}
\tau^{\mu\nu}_{\alpha\beta,\gamma\delta}(k,q)\!\!\!\!&=&\!\!\!-{i\,\kappa\over
2}\left[ \big(I_{\alpha\beta,\gamma\delta}-{1\over
2}\eta_{\alpha\beta}\eta_{\gamma\delta}\big)\left.\bigg[k^\mu
k^\nu+(k-q)^\mu (k-q)^\nu+q^\mu q^\nu-{3\over
2}\eta^{\mu\nu}q^2\right]\right.\nonumber\\
&+&\left.2q_\lambda
q_\sigma\left.\Big[I^{\lambda\sigma,}{}_{\alpha\beta}I^{\mu\nu,}
{}_{\gamma\delta}+I^{\lambda\sigma,}{}_{\gamma\delta}I^{\mu\nu,}
{}_{\alpha\beta}-I^{\lambda\mu,}{}_{\alpha\beta}I^{\sigma\nu,}
{}_{\gamma\delta}-I^{\sigma\nu,}{}_{\alpha\beta}I^{\lambda\mu,}
{}_{\gamma\delta}\right]\right.\nonumber\\
&+&\left.\Big[q_\lambda
q^\mu\big(\eta_{\alpha\beta}I^{\lambda\nu,}{}_{\gamma\delta}
+\eta_{\gamma\delta}I^{\lambda\nu,}{}_{\alpha\beta})+ q_\lambda
q^\nu(\eta_{\alpha\beta}I^{\lambda\mu,}{}_{\gamma\delta}
+\eta_{\gamma\delta}I^{\lambda\mu,}{}_{\alpha\beta}\big)\right.\nonumber\\
&-&\left.q^2\big(\eta_{\alpha\beta}I^{\mu\nu,}{}_{\gamma\delta}+\eta_{\gamma\delta}
I^{\mu\nu,}{}_{\alpha\beta})-\eta^{\mu\nu}q^\lambda
q^\sigma(\eta_{\alpha\beta}
I_{\gamma\delta,\lambda\sigma}+\eta_{\gamma\delta}
I_{\alpha\beta,\lambda\sigma}\big)\Big]\right.\nonumber\\
&+&\left.\Big[2q^\lambda\big(I^{\sigma\nu,}{}_{\gamma\delta}
I_{\alpha\beta,\lambda\sigma}(k-q)^\mu
\!+\!I^{\sigma\mu,}{}_{\gamma\delta}I_{\alpha\beta,\lambda\sigma}(k-q)^\nu\right.\!-\!\left.I^{\sigma\nu,}{}_{\alpha\beta}I_{\gamma\delta,\lambda\sigma}k^\mu\!-\!
I^{\sigma\mu,}{}_{\alpha\beta}I_{\gamma\delta,\lambda\sigma}k^\nu\big)\right.\nonumber\\
&+&\left.q^2\big(I^{\sigma\mu,}{}_{\alpha\beta}I_{\gamma\delta,\sigma}{}^\nu+
I_{\alpha\beta,\sigma}{}^\nu
I^{\sigma\mu,}{}_{\gamma\delta}\big)+\eta^{\mu\nu}q^\lambda q_\sigma
\big(I_{\alpha\beta,\lambda\rho}I^{\rho\sigma,}{}_{\gamma\delta}+
I_{\gamma\delta,\lambda\rho}I^{\rho\sigma,}{}_{\alpha\beta}\big)\Big]\right.\nonumber\\
&+&\left.\Big[\big(k^2+(k-q)^2\big)\left(I^{\sigma\mu,}{}_{\alpha\beta}I_{\gamma\delta,\sigma}{}^\nu
+I^{\sigma\nu,}{}_{\alpha\beta}I_{\gamma\delta,\sigma}{}^\mu-{1\over
2}\eta^{\mu\nu}\big(I_{\alpha\beta,\gamma\delta}-{1\over 2}\eta_{\alpha\beta}\eta_{\gamma\delta})\right.\big)\right.\nonumber\\
&-&\left.\big(k^2\eta_{\alpha\beta}I^{\mu\nu,}{}_{\gamma\delta}+(k-q)^2\eta_{\gamma\delta}
I^{\mu\nu,}{}_{\alpha\beta}\big)\Big]\right.\bigg]\,.
\end{eqnarray}

We work in harmonic (de Donder) gauge which satisfies, in
lowest order,
\begin{equation}
\partial^\mu h_{\mu\nu}={1\over 2}\partial_\nu h\,,\label{eq:dd}
\end{equation}
with
\begin{equation}
h={\rm tr}h_{\mu\nu}\,,
\end{equation}
and in which the graviton propagator has the form
\begin{equation}
D_{\alpha\beta;\gamma\delta}(q)={i\over
q^2+i\epsilon}{1\over 2}(\eta_{\alpha\gamma}\eta_{\beta\delta}+\eta_{\alpha\delta}\eta_{\beta\gamma}
-\eta_{\alpha\beta}\eta_{\gamma\delta})\,.
\end{equation}
Then just as the (massless) photon is described in terms of a
spin-1 polarization vector $\epsilon_\mu$ which can have
projection (helicity) either plus- or minus-1 along the momentum
direction, the (massless) graviton is a spin-2 particle which
can have the projection (helicity) either plus- or minus-2 along
the momentum direction.  Since $h_{\mu\nu}$ is a symmetric tensor,
it can be described in terms of a direct product of unit spin
polarization vectors---
\begin{eqnarray}
{\rm helicity}&=&+2:\quad
h^{(2)}_{\mu\nu}=\epsilon^{+}_\mu \epsilon^{+}_\nu\,,\nonumber\\
{\rm helicity}&=&-2:\quad h^{(-2)}_{\mu\nu}=\epsilon^{-}_\mu
\epsilon^{-}_\nu\,,\label{eq:he}
\end{eqnarray}
and just as in electromagnetism, there is a gauge condition---in
this case Eq.~\eqref{eq:dd}---which must be satisfied. Note that the
helicity states given in Eq.~\eqref{eq:he} are consistent with the
gauge requirement, since
\begin{equation}
\eta^{\mu\nu}\epsilon_\mu^+\epsilon_\nu^+=\eta^{\mu\nu}\epsilon_\mu^-\epsilon_\nu^-=0,\quad{\rm and} \quad
k^\mu\epsilon_\mu^\pm=0\,.
\end{equation}
With this background we can now examine reactions involving gravitons, as discussed in the next section.

\subsection{Graviton photoproduction}\label{sec:gravphotoprod}

\begin{figure}
\begin{center}
\epsfig{file=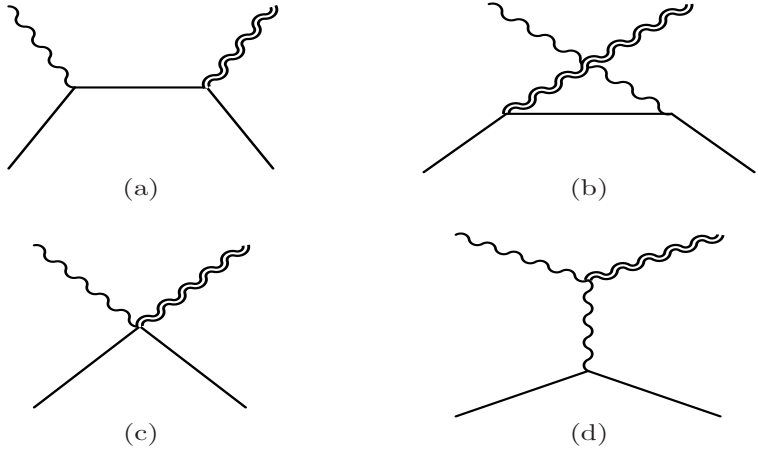,height=6cm,width=10cm} \caption{Diagrams
relevant to graviton photoproduction. }\label{fig:gravphoto}
\end{center}
\end{figure}

We first use the above results to discuss the problem of graviton photoproduction on a target of spin-1---$\gamma+S\rightarrow g+S$---for which the four diagrams we need are shown in Figure~\ref{fig:gravphoto}.  The electromagnetic and gravitational vertices needed for the Born terms and photon pole diagrams---Figures~\ref{fig:gravphoto}a,~\ref{fig:gravphoto}b, and~\ref{fig:gravphoto}d---have been given above.  For the photon pole diagram we require the graviton-photon coupling, which is found from the electromagnetic energy-momentum tensor~\cite{jdj}
\begin{equation}
T_{\mu\nu}=-F_{\mu\alpha}F^\alpha_\nu+{1\over
4}g_{\mu\nu}F_{\alpha\beta}F^{\alpha\beta}\,,
\end{equation}
and yields the photon-graviton vertex\footnote{Note that this form agrees with the previously derived form for the massive graviton-spin-1 energy-momentum tensor---Eq.~\eqref{eq:dt}---in the $m\rightarrow 0$ limit.}
\begin{eqnarray}
\big\langle k_f,\epsilon_f\,\big|V_{grav}^{(\gamma)\mu\nu}\big|\,k_i,\epsilon_i\big\rangle&=&
i\,{\kappa\over 2}\left.\Big[\epsilon_f^*\cdot\epsilon_i\big(k_i^\mu k_f^\nu+k_i^\nu k_f^\mu\big)-\epsilon_f^*\cdot k_i\big(k_f^\mu\epsilon_i^\nu
+\epsilon_i^\mu k_f^\nu\big)\right.\nonumber\\
&-&\left.\epsilon_i\cdot k_f\big(k_i^\nu\epsilon_f^{*\mu}+k_i^\mu\epsilon_f^{*\nu}\big)+k_f\cdot k_i\big(\epsilon_i^\mu\epsilon_f^{*\nu}+\epsilon_i^\nu\epsilon_f^{*\mu}\big)\right.\nonumber\\
&-&\left.\eta^{\mu\nu}\left[k_f\cdot k_i\epsilon_f^*\cdot\epsilon_i-\epsilon_f^*\cdot k_i\epsilon_i\cdot k_f\right]\right.\Big]\,.\label{eq:gs}
\end{eqnarray}

Finally, we need the seagull vertex which arises from the feature that the energy-momentum tensor depends on $p_i,p_f$ and therefore yields a contact interaction when the minimal substitution is made, yielding the spin-1 seagull amplitude shown in Figure~\ref{fig:gravphoto}c.
\begin{eqnarray}
&&\big\langle p_f,\epsilon_B;k_f,\epsilon_f\epsilon_f\,\big|T\big|\,p_i,\epsilon_A;k_i,\epsilon_i\big\rangle_{\rm seagull}=\,{i\over 2}\,\kappa \,e\left.\Big[\epsilon_f^*\cdot(p_f+p_i)\,\epsilon_f^*\cdot\epsilon_i\,\epsilon_B^*\cdot\epsilon_A\,\right.\nonumber\\
&-&\left.\epsilon_B^*\cdot\epsilon_i\,\epsilon_f^*\cdot p_f\,\epsilon_f^*\cdot\epsilon_A-\epsilon_B^*\cdot p_i\,\epsilon_f^*\cdot\epsilon_i\,\epsilon_f^*\cdot\epsilon_A-\epsilon_A\cdot \epsilon_i\,\epsilon_f^*\cdot p_i\,\epsilon_f^*\cdot\epsilon_B^*\right.\nonumber\\
&-&\left.\epsilon_A\cdot p_f\,\epsilon_f^*\cdot\epsilon_i\,\epsilon_f^*\cdot\epsilon_B^*-\epsilon_f^*\cdot\epsilon_A\,\epsilon_i\cdot(p_f+p_i)\,\epsilon_f^*\cdot\epsilon_B^*\right.\Big]\,.
\end{eqnarray}
The individual contributions from the four diagrams in Figure~\ref{fig:gravphoto} are given in Appendix~A and have a rather complex form.  However, when added together we find a {\it much} simpler result---the full graviton photoproduction amplitude is found to be proportional to the already calculated Compton amplitude for spin-1---Eq.~\eqref{eq:gh}---times a universal factor.  That is,
\begin{equation}
\big\langle p_f;k_f,\epsilon_f\epsilon_f\big|T\big|p_i;k_i,\epsilon_i\big\rangle= H\times
\left(\epsilon_{f\alpha}^*\epsilon_{i\beta}
T_{\rm Compton}^{\alpha\beta}(S=1)\right)\,,\label{eq:gi}
\end{equation}
where
\begin{equation}
H={\kappa\over 2e}{p_f\cdot F_f\cdot p_i\over k_i\cdot k_f}=
{\kappa\over 2e}\,{\epsilon_f^*\cdot p_f\,k_f\cdot
p_i-\epsilon_f^*\cdot p_i\,k_f\cdot p_f\over k_i\cdot k_f}\,,
\end{equation}
and $\epsilon_{f\alpha}^*\epsilon_{i\beta} T_{\rm Compton}^{\alpha\beta}(S)$ is the Compton scattering amplitude
for particles of spin-$S$ calculated in the previous section.  The gravitational and electromagnetic
gauge invariance of Eq.~\eqref{eq:gi} is obvious, since it follows directly from the gauge invariance already shown for the Compton amplitude together with the explicit gauge invariance of the factor $H$.  The validity of Eq.~\eqref{eq:gi} allows the calculation of the cross section by helicity methods since the graviton photoproduction helicity amplitudes are given by
\begin{equation}
C^1(ab;cd)=H\times B^1(ab;cd)\,,
\end{equation}
where $B^1(ab;cd)$ are the Compton helicity amplitudes found in the
previous section.  We can then evaluate the invariant photoproduction
cross section using
\begin{equation}
{d\sigma^{\rm photo}_{S=1}\over dt}={1\over 16\pi\big(s-m^2\big)^2}{1\over
3}\sum_{a=-,0,+}{1\over 2}\sum_{c=-,+}\big|C^1(ab;cd)\big|^2\,,\label{eq:ph}
\end{equation}
yielding
\begin{eqnarray}
{d\sigma^{\rm photo}_{S=1}\over dt}&=&-{e^2\kappa^2(m^4-su)\over 96\pi t\big(s-m^2\big)^4\big(u-m^2\big)^2}\left.\Big[(m^4-su+t^2)\big(3(m^4-su)+t^2\big)\right.\nonumber\\
&+&\left.t^2(t-m^2)(t-3m^2)\right.\Big]\,.
\end{eqnarray}
Since
\begin{equation}
|H|={\kappa\over e}\left({m^4-su\over -2t}\right)^{1\over 2}\,,
\end{equation}
the laboratory value of the factor $H$ is
\begin{equation}
|H_{lab}|^2={\kappa^2m^2\over 2e^2}{\cos^2{1\over 2}\theta_L\over
\sin^2{1\over 2}\theta_L}\,,
\end{equation}
the corresponding laboratory cross section is
\begin{eqnarray}
{d\sigma_{{\rm lab},S=1}^{\rm photo}\over
  d\Omega}&=&\big|H_{lab}\big|^2{d\sigma^{\rm Comp}_{{\rm lab},S=1}\over
  dt}\nonumber\\
&=&G\alpha{\omega_f^4\over \omega_i^4}\cos^2{\theta_L\over 2}\left[\bigg({\rm ctn}^2{\theta_L\over 2}\cos^2{\theta_L\over 2}+\sin^2{\theta_L\over 2}\bigg)\bigg(1+2{\omega_i\over m}\sin^2{\theta_L\over 2}\bigg)^2\right.\nonumber\\
&+&\left.{16\omega_i^2\over 3m^2}\sin^2{\theta_L\over 2}\bigg(1+2{\omega_i\over m}\sin^2{\theta_L\over 2}\bigg)+{32\omega_i^4\over 3m^4}\sin^6{\theta_L\over 2}\right]\,.
\end{eqnarray}

The factor $|H_{lab}|^2$ can be thought of as ``dressing" the photon into a graviton.  We see that just as in Compton scattering the low-energy laboratory cross section has a universal form, which is valid for a target of arbitrary spin
\begin{equation}
{d\sigma_{{\rm lab},S}^{\rm photo}\over d\Omega}=G\alpha\cos^2{\theta_L\over 2}\bigg({\rm ctn}^2{\theta_L\over 2}\cos^2{\theta_L\over 2}+\sin^2{\theta_L\over 2}\bigg)\bigg(1+{\cal O}\Big({\omega_i\over m}\Big)\bigg)\,.\label{eq:bz}
\end{equation}
In this case the universality can be understood from the feature that at low energy the leading contribution to the graviton photoproduction amplitude comes {\it not} from the seagull, as in Compton scattering, but rather from the photon pole term,
\begin{equation}
{\rm Amp}_{\gamma-{\rm pole}}\ \ \underset{\omega\ll m}{\longrightarrow}\ \ \kappa{\epsilon_f^*\cdot\epsilon_i\,\epsilon_f^*\cdot k_i\over 2k_f\cdot k_i}\times k_i^\mu\,\big\langle p_f;S,M_f\big|J_\mu\big|p_i;S,M_i\big\rangle\,.
\end{equation}
The leading piece of the electromagnetic current has the universal low-energy structure
\begin{equation}
\big\langle p_f;S,M_f\,\big|J_\mu\big|\,p_i;S,M_i\big\rangle={e\over 2m}\big(p_f+p_i\big)_\mu\delta_{M_f,M_i}\bigg(1+{\cal O}\Big({p_f-p_i\over m}\Big)\bigg)\,,
\end{equation}
where we have divided by the factor $2m$ to account for the normalization of the target particle.  Since $k_i\cdot(p_f+p_i)\underset{\omega\rightarrow 0}{\longrightarrow}2m\omega$, we find the universal low-energy amplitude
\begin{equation}\label{e:NRgpole}
{\rm Amp}_{\gamma-pole}^{NR}=\kappa \,e\, \omega\,{\epsilon_f^*\cdot\epsilon_i\,\epsilon_f^*\cdot k_i\over 2k_f\cdot k_i}\,,
\end{equation}
whereby the resulting helicity amplitudes have the form
\begin{eqnarray}\displaystyle
{\rm Amp}_{\gamma-pole}^{NR}={\kappa \, e\over 2\sqrt{2}}\left\{\begin{array}{ll}
{1\over 2}\sin\theta_L\left.\Big({1+\cos\theta_L\over 1-\cos\theta_L}\right.\Big)={\cos{\theta_L\over 2}\over\sin{\theta_L\over 2}}\cos^2{\theta_L\over 2}& ++=--\,,\\[2ex]
{1\over 2}\sin\theta_L\left.\Big({1-\cos\theta_L\over 1-\cos\theta_L}\right.\Big)={\cos{\theta_L\over 2}\over \sin{\theta\over 2}}\sin^2{\theta_L\over 2}&+-=-+\,.
\end{array}\right.
\end{eqnarray}
Squaring and averaging, summing over initial, final spins we find
\begin{equation}
{d\sigma_{{\rm lab},S}^{\rm photo}\over d\Omega}\ \ \underset{\omega\rightarrow 0}{\longrightarrow}\ \ G\,\alpha\cos^2{\theta_L\over 2}\bigg({\rm ctn}^2{\theta_L\over 2}\cos^2{\theta_L\over 2}+\sin^2{\theta_L\over 2}\bigg)\,.
\end{equation}
as found above---{\it cf.} Eq.~\eqref{eq:bz}.

The power of the factorization theorem is obvious and, as we shall see in the next section, allows the straightforward evaluation of even more complex reactions such as gravitational Compton scattering.\\

\subsection{Gravitational Compton Scattering}\label{sec:gravcomp}

\begin{figure}
\begin{center}
\epsfig{file=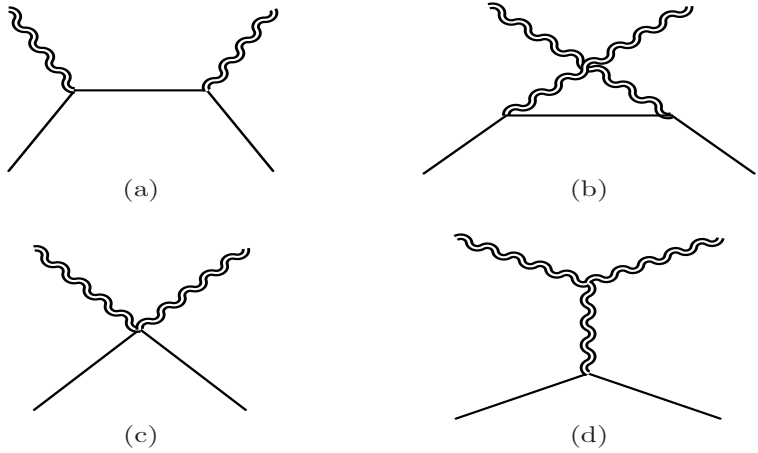,height=6cm,width=10cm} \caption{Diagrams
relevant for gravitational Compton scattering. }\label{fig:gravcomp}
\end{center}
\end{figure}

In the previous section we observed some of the power of the factorization
theorem in the context of graviton photoproduction on a spin-1 target
in that we only needed to calculate the simpler Compton scattering
process rather than to consider the full gravitational interaction.
In this section we tackle a more challenging example, that of
gravitational Compton scattering---$g+S\rightarrow g+S$---from a
spin-1 target, for which there exist the four diagrams shown in
Figure~\ref{fig:gravcomp}.

The contributions from the four individual diagrams can now be calculated and are
quoted in Appendix~A.  Each of the four diagrams has a rather complex
form.  However, when added together the result simplifies
enormously. Defining the kinematic factor
\begin{equation}
Y={\kappa^2\over 8e^4}{p_i\cdot k_i\,p_i\cdot k_f\over k_i\cdot k_f}={\kappa^4\over 16e^4}{(s-m^2)\,(u-m^2)\over t}\,,
\end{equation}
the sum of the four diagrams is found to be given by
\begin{eqnarray}
&&\hskip-2.4cm\big\langle p_f,\epsilon_B;k_f,\epsilon_f\epsilon_f\,\big|{\rm Amp}_{\rm grav}\big|\,p_i,\epsilon_A;k_i,\epsilon_i\epsilon_i\big\rangle_{S=1}\nonumber\\
&=&Y\times]\big\langle p_f,\epsilon_B;k_i,\epsilon_f\,\big|
{\rm Amp}_{\rm em}\big|\,p_i,\epsilon_A;k_i,\epsilon_i\big\rangle_{S=1}
\times\big\langle p_f;k_i,\epsilon_f\,\big|{\rm Amp}_{\rm em}\big|\,p_i;k_i,\epsilon_i\rangle_{S=0}\,,\nonumber\\
\quad\label{eq:cd}
\end{eqnarray}
where
\begin{equation}
\langle p_f;k_i,\epsilon_f|{\rm Amp}_{\rm em}|p_i;k_i,\epsilon_i\rangle_{S=0}=2e^2\left[{\epsilon_i\cdot p_i\,\epsilon_f^*\cdot p_f\over p_i\cdot k_i}-{\epsilon_i\cdot p_f\,\epsilon_f^*\cdot p_i\over p_i\cdot k_f}-\epsilon_f^*\cdot\epsilon_i\right]\,,
\end{equation}
is the Compton amplitude for a spinless target.

In ref. ~\cite{gvp} the identity Eq.~\eqref{eq:cd} was verified for
simpler cases of spin-0 and spin-$\frac12$. This relation is a
consequence of the general relations between gravity and gauge theory
tree-level amplitudes derived from string theory as explained in~\cite{Bjerrum-Bohr:2013bxa}.
Here we have shown its validity for the much more complex case of spin-1 scattering.  The corresponding cross section can be calculated by helicity methods using the identity
\begin{equation}
D^1(ab;cd)=Y\times B^1(ab;cd)\times A^0(cd)\,,
\end{equation}
where $B^1(ab;cd)$ is the spin-1 Compton helicity amplitude calculated in section~\ref{sec:review} while
\begin{eqnarray}
A^0(++)&=&2e^2{m^4-su\over \big(s-m^2)\big(u-m^2)}\,,\nonumber\\
A^0(+-)&=&2e^2{-m^2t\over \big(s-m^2)\big(u-m^2)}\,,
\end{eqnarray}
are the helicity amplitudes for spin zero Compton scattering.  Using Eq.~\eqref{eq:cd} the invariant cross section for unpolarized spin-1 gravitational Compton scattering
\begin{equation}
{d\sigma^{\rm g-Comp}_{S=1}\over dt}={1\over 16\pi\big(s-m^2\big)^2}\ { 1\over
3}\sum_{a=-,0,+}{1\over 2}\sum_{c=-,+}\big|D^1(ab;cd)\big|^2\,,\label{eq:mh}
\end{equation}
is found to be
\begin{eqnarray}
{d\sigma^{\rm g-Comp}_{S=1}\over dt}&=&{\kappa^4\over 768\pi\big(s-m^2\big)^4\big(u-m^2\big)^2t^2}
\left[(m^4-su)^2\big(3(m^4-su)+t^2)(m^4-su+t^2)\big)\right.\nonumber\\
&+&\left.m^4t^4(3m^2-t)(m^2-t)\right]\,.
\end{eqnarray}
This form can be compared with the corresponding unpolarized gravitational Compton cross sections found in ref. ~\cite{gvp}
\begin{eqnarray}
{d\sigma^{\rm g-Comp}_{S={1\over 2}}\over dt}&=&{\kappa^4\over 512\pi}{\left(\big(m^4-su\big)^3\big(2(m^4-su)+t^2\big)+m^6t^4\big(2m^2-t\big)\right)\over t^2\big(s-m^2\big)^4\big(u-m^2\big)^2}\,,\nonumber\\
{d\sigma^{\rm g-Comp}_{S=0}\over d\Omega}&=&{\kappa^4\over 256\pi^2\big(s-m^2\big)^4\big(u-m^2\big)^2t^2}\left.\Big[(m^4-su)^4+m^8t^4\right.\Big]\,.
\end{eqnarray}
The corresponding laboratory frame cross sections are
\begin{eqnarray}
{d\sigma_{{\rm lab},S=1}^{\rm g-Comp}\over d\Omega}&=&G^2m^2{\omega_f^4\over \omega_i^4}\left[\bigg({\rm ctn}^4{\theta_L\over 2}\cos^4{\theta_L\over 2}+\sin^4{\theta_L\over 2}\bigg)\bigg(1+2{\omega_i\over m}\sin^2{\theta_L\over 2}\bigg)^2\right.\nonumber\\
&+&\left.{16\over 3}{\omega_i^2\over m^2}\bigg(\cos^6{\theta_L\over 2}+\sin^6{\theta_L\over 2}\bigg)\bigg(1+2{\omega_i\over m}\sin^2{\theta_L\over 2}\bigg)\right.\nonumber\\
&+&\left.{16\over 3}{\omega_i^4\over m^4}\sin^2{\theta_L\over 2}\bigg(\cos^4{\theta_L\over 2}+\sin^4{\theta_L\over 2}\bigg)\right]\,,\nonumber\\
{d\sigma^{\rm g-Comp}_{{\rm lab},S={1\over 2}}\over d\Omega}&=&G^2m^2{\omega_f^3\over \omega_i^3}\left[\bigg({\rm ctn}^4{\theta_L\over 2}\cos^4{\theta_L\over 2}+\sin^4{\theta_L\over 2}\bigg)+2{\omega_i\over m}\bigg({\rm ctn}^2{\theta_L\over 2}\cos^6{\theta_L\over 2}+\sin^6{\theta_L\over 2}\bigg)\right.\nonumber\\
&+&\left.2{\omega_i^2\over m^2}\bigg(\cos^6{\theta_L\over 2}+\sin^{6}{\theta_L\over 2}\bigg)\right]\,,\nonumber\\
{d\sigma^{\rm g-Comp}_{{\rm lab},S=0}\over d\Omega}&=&G^2m^2{\omega_f^2\over\omega_i^2}\left[{\rm ctn}^4{\theta_L\over 2}\cos^4{\theta_L\over 2}+\sin^4{\theta_L\over 2}\right]\,.\emph{}
\end{eqnarray}
We observe that the low-energy laboratory cross section has the universal form for any spin
\begin{equation}
{d\sigma_{{\rm lab},S}^{\rm g-Comp}\over d\Omega}=G^2m^2\left[{\rm ctn}^4{\theta_L\over 2}\cos^4{\theta_L\over 2}+\sin^4{\theta_L\over 2}+{\cal O}\Big({\omega_i\over m}\Big)\right]\,.\label{eq:zs}
\end{equation}

It is interesting to note that the ``dressing" factor for the leading (++) helicity Compton amplitude---
\begin{equation}
\big |Y\big|\,\big|A^{++}\big|\ =\ {\kappa^2\over 2e^2}\ {m^4-su\over -t}\ \ \overset{\rm lab}{\longrightarrow}\ \ {\kappa^2m^2\over 2e^2}\ {\cos^2{\theta_l\over 2}\over \sin^2{\theta\over 2}}\,,
\end{equation}
---is simply the square of the photoproduction dressing factor $H$, as might intuitively be expected since now {\it both} photons must be dressed in going from the Compton to the gravitational Compton cross section.\footnote{In the case of $+-$ helicity the ``dressing" factor is
\begin{equation}
\big|Y\big|\,\big|A^{+-}\big|={\kappa^2\over 2e^2}\ m^2\,.
\end{equation}
so that the nonleading contributions will have different dressing factors.}
In this case the universality of the nonrelativistic cross section follows from the leading contribution arising from the graviton pole term
\begin{equation}
{\rm Amp}_{g-pole}\ \ \underset{\omega\ll m}{\longrightarrow}\ \
{\kappa\over 4k_f\cdot k_i}\ \big(\epsilon_f^*\cdot\epsilon_i\big)^2\ \big(k_f^\mu k_f^\nu+k_i^\mu k_i^\nu\big)\ {\kappa\over 2}\ \big \langle p_f;S,M_f\,\big|T_{\mu\nu}\big |\,p_i;S,M_i\big\rangle\,.
\end{equation}
Here the matrix element of the energy-momentum tensor has the universal low-energy structure
\begin{equation}
{\kappa\over 2}\,\big \langle p_f;S,M_f\,\big|T_{\mu\nu}\big |\,p_i;S,M_i\big\rangle={\kappa\over 4m}\,\big(p_{f\mu}p_{i\nu}+p_{f\nu}p_{i\mu}\big)\,\delta_{M_f,M_i}\Big(1+{\cal O}\Big({p_f-p_i\over m}\Big)\Big)\,,
\end{equation}
where we have divided by the factor $2m$ to account for the normalization of the target particle.  We find then the universal form for the leading graviton pole amplitude
\begin{equation}
{\rm Amp}_{g-pole}\ \ \underset{\rm non-rel}{\longrightarrow}\ \
{\kappa^2\over 8m\, k_f\cdot k_i}\ \big(\epsilon_f^*\cdot\epsilon_i\big)^2\,\big(p_i\cdot k_f\,p_f\cdot k_f+p_i\cdot k_i\, p_f\cdot k_i\big)\,\delta_{M_f,M_i}\,.
\end{equation}
Since $p\cdot k\underset{\omega\ll m}{\longrightarrow}m\omega$ the corresponding helicity amplitudes become
\begin{eqnarray}
{\rm Amp}_{g-pole}^{\rm NR}=4\pi Gm\left\{\begin{array}{ll}
{\big(1+\cos\theta_L\big)^2\over 2\big(1-\cos\theta_L\big)}={\cos^4{\theta_L\over 2}\over \sin^2{\theta_L\over 2}}& ++=--\,,\\[2ex]
{\big(1-\cos\theta_L\big)^2\over 2\big(1-\cos\theta_L\big)}={\sin^4{\theta_L\over 2}\over \sin^2{\theta_L\over 2}}&+-=-+\,.
\end{array}\right.
\end{eqnarray}
Squaring and averaging, summing over initial, final spins we find
\begin{equation}
{d\sigma_{{\rm lab},S}^{\rm g-Comp}\over d\Omega}\ \ \underset{\omega\rightarrow 0}{\longrightarrow}\ \ G^2\,m^2\,\bigg[{\rm ctn}^4{\theta_L\over 2}\cos^4{\theta_L\over 2}+\sin^4{\theta_L\over 2}\bigg]\,,
\end{equation}
as found in Eq.~\eqref{eq:zs} above.

\section{Graviton-Photon Scattering}\label{sec:gravpho}
In the previous sections we have generalized the results of reference~\cite{gvp} to the case of a massive spin-1 target.  Here we show how these techniques can be used to calculate the cross section for photon-graviton scattering.  In the Compton scattering calculation we assumed that the spin-1 target had charge $e$.  However, the photon couplings to the graviton are identical to those of a graviton coupled to a charged spin-1 system in the massless limit, and one might assume then that, since the results of the gravitational Compton scattering are independent of charge, the graviton-photon cross section can be calculated by simply taking the $m\rightarrow 0$ limit of the graviton-spin-1 cross section.  Of course, the laboratory cross section no longer makes sense since the photon cannot be brought to rest, but the invariant cross section is well defined in this limit---
\begin{equation}
{d\sigma_{S=1}^{\rm g-Comp}\over dt}\ \ \underset{m\rightarrow 0}{\longrightarrow}\ \ {4\pi \,G^2\,\big(3s^2u^2-4t^2su+t^4\big)\over 3s^2t^2}\,,\label{eq:hd}
\end{equation}
and it might be naively assumed that Eq.~\eqref{eq:hd} is the graviton-photon scattering cross section.  However, this is {\it not} the case and the resolution of this problem involves some interesting physics.

We begin by noting that in the massless limit the only nonvanishing helicity amplitudes are
\begin{eqnarray}
D^1(++;++)_{m=0}&=&D^1(--;--)_{m=0}=8\pi\, G\,{s^2\over t}\,,\nonumber\\
D^1(--;++)_{m=0}&=&D^1(++;--)_{m=0}=8\pi\, G\,{u^2\over t}\,,\nonumber\\
D^1(00;++)_{m=0}&=&D^1(00;--)_{m=0} \ \,=8\pi\, G\,{su\over t}\,,
\end{eqnarray}
which lead to the cross section
\begin{eqnarray}
{d\sigma^{\rm g-Comp}_{S=1}\over dt}&=& {1\over 16\pi s^2}\,\frac13\,\sum_{a=+,0,-}\frac12\sum_{c=+,-}\big|D^1(ab;cd)\big|^2\cr
&=&{1\over 16\pi s^2}\,{1\over 3\cdot 2}\,(8\pi G)^2\times
2\times\left[{s^4\over t^2}+{u^4\over t^2}+{s^2u^2\over t^2}\right]\cr
&=&{4\pi\over 3}\,G^2{s^4+u^4+s^2u^2\over s^2t^2}\,,
\end{eqnarray}
in agreement with Eq.~\eqref{eq:hd}.  However, this result demonstrates the problem.  We know that in Coulomb gauge the photon has only two transverse degrees of freedom, corresponding to positive and negative helicity---there exists {\it no} longitudinal degree of freedom.  Thus the correct photon-graviton cross section is actually
\begin{eqnarray}
{d\sigma_{g\gamma}\over dt}&=& {1\over 16\pi s^2}\,\frac13\sum_{a=+,-}\frac12\sum_{c=+,-}\big|D^1(ab;cd)\big|^2\cr
&=&{1\over 16\pi s^2}\,{1\over 2\cdot 2}\,\big(8\pi G\big)^2\times 2\times\left[{s^4\over t^2}+{u^4\over t^2}\right]=2\pi\, G^2\,{s^4+u^4\over s^2t^2}\,,
\end{eqnarray}
which agrees with the value calculated via conventional methods by Skobelev~\cite{skb}.  Alternatively, since in the center of mass frame
\begin{equation}
{dt\over d\Omega}={\omega_{\rm CM}\over \pi}\,,
\end{equation}
we can write the center of mass graviton-photon cross section in the form
\begin{equation}\label{e:gG}
{d\sigma_{\rm CM}\over d\Omega}=2\,G^2\,\omega_{\rm CM}^2\left({1+\cos^8{\theta_{\rm CM}\over 2}\over \sin^4{\theta_{\rm CM}\over 2}}\right)\,,
\end{equation}
again in agreement with the value given by Skobelev~\cite{skb}.

So what has gone wrong here?  Ordinarily in the massless limit of a spin-1 system, the longitudinal mode decouples because the zero helicity spin-1 polarization vector becomes
\begin{equation}
\epsilon_\mu^0\ \ \underset{m\rightarrow 0}{\longrightarrow}\ \ {1\over m}\bigg(p,\Big(p+{m^2\over 2p}+\ldots\Big)\hat{z}\bigg)={1\over m}p_\mu+\Big(0,{m\over 2p}\hat{z}\Big)+\ldots\label{eq:fc}
\end{equation}
However, the term proportional to $p_\mu$ vanishes when contracted with a conserved current by gauge invariance while the term in ${m\over 2p}$ vanishes in the massless limit.  That the spin-1 Compton scattering amplitude becomes gauge invariant for the spin-1 particles in the massless limit can be seen from the fact that the Compton amplitude can be written as
\begin{eqnarray}
&&\hskip-1.4cm{\rm Amp}^{\rm Comp}_{S=1}\underset{m\rightarrow 0}{\longrightarrow}{e^2\over p_i\cdot q_i\,p_i\cdot q_f}\left.\bigg[{\rm Tr}\big(F_iF_fF_AF_B\big)+{\rm Tr}\big(F_iF_AF_fF_B\big)+{\rm Tr}\big(F_iF_AF_BF_f\big)\right.\nonumber\\
&&\hskip1.5cm-\left.{1\over 4}\left.\Big({\rm Tr}\big(F_iF_f\big){\rm Tr}\big(F_AF_B\big)
+{\rm Tr}\big(F_iF_A\big){\rm Tr}\big(F_fF_B\big)+{\rm Tr}\big(F_iF_B\big){\rm Tr}\big(F_fF_A\big)\right.\Big)\right]\,,\nonumber\\
\quad
\end{eqnarray}
which can be checked by a bit of algebra.  Equivalently, one can verify
that the massless spin-1 amplitude vanishes if one replaces either
$\epsilon_{A\mu}$ by $p_{i\mu}$ or $\epsilon_{B\mu}$ by $p_{f\mu}$.
However, what happens when we have {\it two} longitudinal spin-1
particles is that the product of longitudinal polarization vectors is
proportional to $1/m^2$, while the correction term to the
four-momentum $p_\mu$ is ${\cal O}(m^2)$ so that the product is
nonvanishing in the massless limit.  That is why the multipole
$D(00;++)_{m=0}=D(00;--)_{m=0}$ is nonzero.  One can deal with this problem by simply omitting the longitudinal degree of freedom explicitly, as we did above, but this seems a rather crude way to proceed.  Should not this behavior arise naturally?

The problem here is that as long as the mass of the spin-1 particle remains finite everything is fine.  However, when the spin-1 particle becomes massless the theory becomes undefined.  This can be seen from the neutral spin-1 (Proca) Lagrangian, which has the form
\begin{equation}
{\cal L}^1=-{1\over 4}F_{\mu\nu}F^{\mu\nu}+{1\over 2}m^2A_\mu A^\mu=-{1\over 2}\big(\partial_\mu A_\nu\partial^\mu A^\nu-\partial_\mu A_\nu\partial^\nu A^\mu\big)+{1\over 2}m^2A_\mu A^\mu\,.\label{eq:vx}
\end{equation}
The classical equation of motion then becomes
\begin{equation}
\partial^\mu F_{\mu\nu}+m^2 A_\nu=0\label{eq:sx}\,.
\end{equation}
Taking the divergence of Eq.~\eqref{eq:sx} we find
\begin{equation}
m^2\partial^\nu A_\nu=0\,,
\end{equation}
which yields the constraint $m^2\partial^\nu A_\nu=0$.  Then provided that $m^2\neq 0$ we have the stricture $\partial^\nu A_\nu=0$, which is the condition that changes the number of degrees of freedom from four to three, as required for a spin-1 particle.  However, in the massless limit, this is no longer the case.  Another way to see this is to integrate by parts, whereby Eq.~\eqref{eq:vx} can be written in the form
  \begin{equation}
{\cal L}^1_{m=0}={1\over 2}A_\mu {\cal O}^{\mu\nu} A_\nu\,,\quad{\rm with}\quad {\cal O}^{\mu\nu}=\eta^{\mu\nu}\Box-\partial^\mu\partial^\nu\,.
\end{equation}
In particle physics the photon propagator is given by the inverse of this operator---${\cal O}^{-1}_{\mu\nu}$---which is defined via ${\cal O}^{\mu\nu}{\cal O}^{-1}_{\nu\alpha}=\delta^\mu_\alpha$~\cite{bhb}.  However, the operator ${\cal O}^{\mu\nu}$ does not have an inverse, since it has a zero eigenvalue, as can be seen by operating on a quantity of the form $\partial_\nu \Lambda(x)$ where $\Lambda(x)$ is an arbitrary scalar function.  The solution to this problem is well known.  The Lagrangian must be altered by adding a gauge fixing term
\begin{equation}
{\cal L}_{m=0}^1\longrightarrow -{1\over 4}F_{\mu\nu}-{\lambda\over 2}\big(\partial_\mu A^\mu\big)^2\,,
\end{equation}
where $\lambda$ is an arbitrary constant.  We now have ${\cal O}^{\mu\nu}=\eta^{\mu\nu}\Box-(1-\lambda)\partial^\mu\partial^\nu$ which {\it does} possess an inverse---${\cal O}^{-1}_{\mu\nu}={1\over \Box}\left(\eta_{\mu\nu}-{1-\lambda\over \lambda}{\partial_\mu\partial_\nu\over \Box}\right)$. It is this gauge fixing term, which is required in the massless limit, and which eliminates the longitudinal degree of freedom.  This degree of freedom acts like simple scalar field (spin-0 particle) and must be subtracted from the massless limit of the spin-1 result.  Indeed, from ref.~\cite{gvp} we see that the massless limit of the $++$ graviton scattering from a spin-0 target becomes
\begin{equation}
D^0(++)=(2e^2)^2\times Y=8\pi G{su\over t}\,,
\end{equation}
while the $+-$ helicity amplitude vanishes.  This scalar amplitude is identical to the amplitude $D_1(00;++)$ and eliminates the longitudinal degree of freedom when subtracted from the massless spin-1 limit.

An alternative way to obtain this result is to use the Stueckelberg form of the spin-1 Lagrangian, which involves coupling a new spin-0 field $B$~\cite{stb}
\begin{equation}
{\cal L}_S=-{1\over 4}F_{\mu\nu}F^{\mu\nu}+{m^2\over 2}\Big(A_\mu+{1\over m}\partial_\mu B\Big)\Big(A^\mu+{1\over m}\partial^\mu B\Big)-{1\over 2}\big(\partial_\mu A^\mu+mB\big)\big(\partial_\nu A^\nu+mB\big)\,.\label{eq:xs}
\end{equation}
As long as $m\neq 0$ the fields $A_\mu$ and $B$ are coupled.  However, if we take the massless limit Eq.~\eqref{eq:xs} becomes
\begin{equation}
{\cal L}_S\underset{m\rightarrow 0}{\longrightarrow} -{1\over 4}F_{\mu\nu}F^{\mu\nu}-{1\over 2}\partial_\mu A^\mu\partial_\nu A^\nu+{1\over 2}\partial_\mu B\partial^\mu B\,,
\end{equation}
and represents the sum of two independent massless fields---a spin-1
component $A_\mu$ with the Lagrangian (in Feynman gauge $\lambda=1$)
\begin{equation}
{\cal L}_S^1=-{1\over 4}F_{\mu\nu}F^{\mu\nu}-{1\over 2}\partial_\mu A^\mu\partial_\nu A^\nu=-{1\over 2}A^\mu\Box A_\mu\,,
\end{equation}
for which we {\it do} have an inverse and an independent spin-0 component having the Lagrangian
\begin{equation}
{\cal L}_S^0={1\over 2}\partial_\mu B\partial^\mu B\,.
\end{equation}
It is the scattering due to the spin-1 component which is physical and leads to the graviton-photon scattering amplitude, while the spin-0 component is {\it unphysical} and generates the longitudinal component of the massless limit of the graviton-spin-1 scattering.

As a final comment we note that the graviton-graviton scattering amplitude can be obtained by dressing the product of two massless spin-1 Compton amplitudes~\cite{str}---
\begin{eqnarray}
\big\langle p_f,\epsilon_B\epsilon_B;k_f,\epsilon_f\epsilon_f\,\big|{\rm Amp}^{\rm tot}_{grav}\big|\,p_i\epsilon_A\epsilon_A;k_i,\epsilon_i\epsilon_i\big\rangle_{m=0,S=2}&&\nonumber\\
&&\hskip-3cm=Y\times\langle p_f,\epsilon_B;k_f,\epsilon_f\,|{\rm Amp}^{\rm Comp}_{em}|\,p_i,\epsilon_A;k_i\epsilon_i\rangle_{m=0,S=1}\nonumber\\
&&\hskip-3cm\times\langle p_f,\epsilon_B;k_f,\epsilon_f\,|{\rm Amp}^{\rm Comp}_{em}|\,p_i,\epsilon_A;k_i\epsilon_i\rangle_{m=0,S=1}\,.
\end{eqnarray}
Then for the helicity amplitudes we have
\begin{equation}
E^2(++;++)_{m=0}=Y\big(B^1(++;++)_{m=0}\big)^2\,,
\end{equation}
where $E^2(++;++)$ is the graviton-graviton $++;++$ helicity amplitude while $B^1(++;++)$ is the corresponding spin-1 Compton helicity amplitude.  Thus we find
\begin{equation}
E^2(++;++)_{m=0}={\kappa^2\over 16e^4}\,{su\over t}\times \bigg({2e^2\,{s\over u}}\bigg)^2=8\pi \,G\,{s^3\over ut}\,,
\end{equation}
which agrees with the result calculated via conventional methods~\cite{pfg}.  In this case there exist non zero helicity amplitudes related by crossing symmetry.  However, we defer detailed discussion of this result to a future communication.

\section{The forward cross section}\label{sec:forward}

The forward limit, {\it i.e.}, $\theta_L\to 0$,  of the laboratory frame,
Compton cross sections evaluated in section~\ref{sec:compton}
has a universal structure independent of the spin $S$ of the massive target
\begin{equation}\label{e:sigmaCompton}
\lim_{\theta_L\to 0}   {d\sigma^{\rm Comp}_{{\rm lab},S}\over d\Omega} ={\alpha^2\over2m^2}\,,
\end{equation}
reproducing the Thomson scattering cross section.

\medskip
For graviton photoproduction, the small angle limit is very different,
since the forward scattering cross section is divergent---the small angle limit of the graviton photoproduction of
section~\ref{sec:gravphotoprod} is given by

\begin{equation} \label{e:fg}
  \lim_{\theta_L\to0}   {d\sigma^{\rm photo}_{{\rm lab},S}\over d\Omega} = {4G\alpha\over \theta_L^2}\,,
\end{equation}
and arises from the photon pole in
figure~\ref{fig:gravphoto}(d). Notice that this behavior differs from
the familiar $1/\theta^4$ small-angle Rutherford cross section for
scattering in a Coulomb-like potential. This divergence of the forward
cross section indicates that a long range force is involved but
with an effective $1/r^2$ potential. This effective potential arising
from the $\gamma$-pole in figure~\ref{fig:gravphoto}(d), is the
Fourier transform with respect to the momentum transfer $q=k_f-k_i$ of the low-energy limit given in
Eq.~\eqref{e:NRgpole}. Because of the linear dependence in the momenta in the
numerator one obtains
\begin{equation}
\int {d^3\vec q\over (2\pi)^3}\,  e^{i \vec q\cdot \vec r} \,
  {1\over |\vec q|} = {1\over 2\pi^2 r^2}\,,
\end{equation}
and this leads to the peculiar forward scattering behavior of the cross section.
Another contrasting feature of graviton photoproduction is the
independence of the forward cross section on the mass $m$ of the target.

\medskip

The small angle limit of the gravitational Compton
cross section derived in section~\ref{sec:gravcomp} is given by

\begin{equation}\label{e:fG}
  \lim_{\theta_L\to0}
  {d\sigma^{\rm g-Comp}_{lab,S}\over d\Omega} = {16G^2m^2\over \theta_L^4}\,.
\end{equation}
The limit is, of course, independent of the spin $S$ of the matter field.
Finally, the photon-graviton cross section derived in
section~\ref{sec:gravpho}, has the forward scattering dependence
\begin{equation}\label{e:fG2}
  \lim_{\theta_{\rm CM}\to0} {d\sigma_{\rm CM}\over d\Omega}=
  {32G^2\omega_{\rm CM}^2\over \theta_{\rm CM}^4}  \,.
\end{equation}
The behaviors in Eq.~\eqref{e:fG} and~\eqref{e:fG2} are due to the graviton pole in
figure~\ref{fig:gravcomp}(d), and are typical of the small-angle behavior of Rutherford scattering in a Coulomb potential.

The classical bending of the geodesic for a
massless particle in a Schwarzschild metric produced by a point-like mass $m$ is given by $b=4Gm/\theta+O(1)$~\cite{Landau:1982dva}, where $b$ is the classical impact parameter.  The associated classical cross section is
\begin{equation}
{  d\sigma^{\rm classical}  \over d\Omega}= {b\over \sin\theta}
\left|db\over d\theta\right| \simeq {16G^2m^2\over \theta^4}+O(\theta^{-3})\,,
\end{equation}
matching the expression in Eq.~\eqref{e:fG}.
The diagram in figure~\ref{fig:gravcomp}(d) describes the
gravitational interaction between a massive particle of spin-$S$ and a
graviton. In the forward scattering limit the remaining diagrams of
figure~\ref{fig:gravcomp} have vanishing contributions.
Since this limit is independent of the spin of the particles interacting
gravitationally, the expression in Eq.~\eqref{e:fG}
describes the forward gravitational scattering cross section of \emph{any} massless
particle on the target of mass $m$ and explains the match with the
classical formula given above.

Eq.~\eqref{e:fG2} can be interpreted in a similar way, as the bending
of a geodesic in a geometry curved by the energy density with an
effective Schwarzschild radius of $\sqrt2\,G\omega_{\rm CM}$ determined by the
center-of-mass energy~\cite{Amati:1987uf}.
However, the effect is fantastically small since the
cross section in Eq.~\eqref{e:fG2} is of order
$\ell_P^4/(\lambda^2\, \theta_{\rm CM}^4)$ where $\ell_P^2=\hbar G/c^3\sim1.62\,10^{-35}\,$m
is the Planck length, and $\lambda$ the wavelength of the photon.

\section{Conclusion}\label{sec:conclusion}

In ref. ~\cite{gvp} it was demonstrated that the gravitational
interactions of a charged spin-0 or spin-$\frac12$ particle are
greatly simplified by use of the recently discovered factorization
theorem, which asserts that the gravitational amplitudes must be
identical to corresponding electromagnetic amplitudes multiplied by
universal kinematic factors.  In the present work we demonstrated that
the same simplification applies when the target particle carries
spin-1. Specifically, we evaluated the graviton photoproduction and
graviton Compton scattering amplitudes explicitly using direct and
factorized techniques and showed that they are identical.  However,
the factorization methods are {\it enormously} simpler and allow the
use of familiar electromagnetic calculational methods, eliminating the
need for the use of less familiar and more cumbersome tensor
quantities.  We also studied the massless limit of the spin-1 system
and showed how the use of factorization permits a relatively simple
calculation of graviton-photon scattering.  Finally, we discussed a
subtlety in this graviton-photon calculation having to do with the
feature that the spin-1 system must change from three to two degrees
of freedom when $m\rightarrow 0$ and studied why the zero mass limit
of the spin-1 gravitational Compton scattering amplitude does not
correspond to that for photon scattering.  We noted that graviton-graviton
scattering is also simply obtained by taking the product of Compton
amplitudes dressed by the appropriate kinematic factor.

We discussed the main feature of the forward cross section for each
process studied in this paper. Both the Compton and the gravitational
Compton scattering have the expected behavior, while graviton
photoproduction has a different shape that could in principle lead to an
interesting new experimental signature of a graviton scattering on
matter. An extension of the present discussion at loop
order and implications for the photoproduction of gravitons from
stars~\cite{DeLogi:1977qe,Papini:1977fm} will be given elsewhere.

 \section*{ACKNOWLEDGMENTS}
We would like to thank Thibault Damour, Poul H. Damgaard, John Donoghue and Gabriele
Veneziano for comments and discussions. The research of P.V. was supported by the ANR grant   reference QST
ANR 12 BS05 003   01, and the CNRS grants PICS number 6076 and 6430.\\ \\

\begin{center}
{\bf APPENDIX~A}
\end{center}
Here we give the detailed contributions from each of the four diagrams contributing to graviton photoproduction and to gravitational Compton scattering.  In the case of graviton photoproduction---Figure~\ref{fig:gravphoto}---we have the four pieces
\\ \\

\begin{center}
{\bf Graviton photoproduction: spin-1}
\end{center}
\begin{eqnarray}\!\!\!\!\!\!\!\!\!\!\!\!\!\!\!\!
{\rm Born-a}:&& \!\!\!\!\!\!\!\!\!{{\rm Amp}_a(S=1)}={\kappa e\over p_i\cdot k_i}\,\left.\bigg[\epsilon_i\cdot p_i\left[\epsilon_B^*\cdot\epsilon_A\,\epsilon_f^*\cdot p_f\,\epsilon_f^*\cdot p_f-\epsilon_B^*\cdot k_f\,\epsilon_f^*\cdot p_f\,\epsilon_f^*\cdot\epsilon_A\right.\right.\nonumber\\
&-&\left.\left.\epsilon_A\cdot p_f\,\epsilon_f^*\cdot p_f\,\epsilon_f^*\cdot\epsilon_B^*+p_f\cdot k_f\,\epsilon_f^*\cdot\epsilon_A\,\epsilon_f^*\cdot\epsilon_B^*\right]\right.\nonumber\\
&+&\left.\epsilon_A\cdot\epsilon_i\left[\epsilon_B^*\cdot k_i\,\epsilon_f^*\cdot p_f\,\epsilon_f^*\cdot p_f-\epsilon_B^*\cdot k_f\,\epsilon_f^*\cdot p_f\,\epsilon_f^*\cdot k_i-p_f\cdot k_i\,\epsilon_f^*\cdot p_f\,\epsilon_f^*\cdot\epsilon_B^*\right.\right.\nonumber\\
&+&\left.\left.p_f\cdot k_f\,\epsilon_f^*\cdot k_i\,\epsilon_f^*\cdot\epsilon_B^*\right]\right.\nonumber\\
&-&\left.\epsilon_A\cdot k_i\,\left[\epsilon_B^*\cdot\epsilon_i\,\epsilon_f^*\cdot p_f\,\epsilon_f^*\cdot p_f-\epsilon_B^*\cdot k_f\,\epsilon_f^*\cdot p_f\,\epsilon_f^*\cdot\epsilon_i-\epsilon_i\cdot p_f\,\epsilon_f^*\cdot p_f\,\epsilon_f^*\cdot\epsilon_B^*\right.\right.\nonumber\\
&+&\left.\left.p_f\cdot k_f\,\epsilon_f^*\cdot\epsilon_i\,\epsilon_f^*\cdot\epsilon_B^*\right]\right.\nonumber\\
&-&\left.\epsilon_B^*\cdot\epsilon_f^*\,\epsilon_A\cdot\epsilon_i\,\epsilon_f^*\cdot
  p_fp_i\cdot k_i\right.\bigg]\,.
\end{eqnarray}
\begin{eqnarray}\!\!\!\!\!\!\!\!\!\!\!\!\!\!\!\!
{\rm Born-b}:&& \!\!\!\!\!\!\!\!\!{{\rm Amp}_b(S=1)}=-{\kappa e\over p_i\cdot k_f}\,\left.\bigg[\epsilon_i\cdot p_f\left[\epsilon_A\cdot\epsilon_B^*\,\epsilon_f^*\cdot p_i\,\epsilon_f^*\cdot p_i-\epsilon_B^*\cdot p_i\,\epsilon_f^*\cdot p_i\,\epsilon_f^*\cdot\epsilon_A\right.\right.\nonumber\\
&+&\left.\left.\epsilon_A\cdot k_f\,\epsilon_f^*\cdot p_i\,\epsilon_f^*\cdot\epsilon_B^*-p_i\cdot k_f\,\epsilon_f^*\cdot\epsilon_A\,\epsilon_f^*\cdot\epsilon_B^*\right]\right.\nonumber\\
&+&\left.\epsilon_B^*\cdot k_i\left[\epsilon_A\cdot\epsilon_i\,\epsilon_f^*\cdot p_i\,\epsilon_f^*\cdot p_i-\epsilon_i\cdot p_i\,\epsilon_f^*\cdot p_i\,\epsilon_f^*\cdot\epsilon_A+\epsilon_A\cdot k_f\,\epsilon_f^*\cdot p_i\,\epsilon_f^*\cdot\epsilon_i\right.\right.\nonumber\\
&-&\left.\left.p_i\cdot k_f\,\epsilon_f^*\cdot\epsilon_A\,\epsilon_f^*\cdot\epsilon_i\right]\right.\nonumber\\
&+&\left.\epsilon_i\cdot\epsilon_B^*\left[\epsilon_A\cdot k_i\,\epsilon_f^*\cdot p_i\,\epsilon_f^*\cdot p_i-p_i\cdot k_i\,\epsilon_f^*\cdot p_i\,\epsilon_f^*\cdot\epsilon_A+\epsilon_A\cdot k_f\,\epsilon_f^*\cdot p_i\,\epsilon_f^*\cdot k_i\right.\right.\nonumber\\
&-&\left.\left.p_i\cdot k_f\,\epsilon_f^*\cdot\epsilon_A\,\epsilon_f^*\cdot k_i\right]\right.\nonumber\\
&-&\left.\epsilon_A\cdot\epsilon_f^*\,\epsilon_f^*\cdot p_i\,\epsilon_B^*\cdot\epsilon_i\,p_i\cdot k_f\right.\bigg]\,.\nonumber\\
\end{eqnarray}
\begin{eqnarray}
{\rm Seagull-c}:&&\!\!\!\!\!\!\!\!\!{\rm Amp}_c(S=1)=\kappa \,e\left.\bigg[\epsilon_f^*\cdot\epsilon_i(\epsilon_B^*\cdot\epsilon_A\,\epsilon_f^*\cdot (p_f+p_i)-\epsilon_A\cdot p_f\,\epsilon_B^*\cdot\epsilon_f^*
-\epsilon_B^*\cdot p_i\,\epsilon_A\cdot\epsilon_f^*)\right.\nonumber\\
&-&\left.\epsilon_B^*\cdot\epsilon_f^*\,\epsilon_A\cdot\epsilon_i\,\epsilon_f^*\cdot p_i-\epsilon_A\cdot\epsilon_f^*\,\epsilon_B^*\cdot\epsilon_i\,\epsilon_f^*\cdot p_f+\epsilon_f^*\cdot\epsilon_A\,\epsilon_f^*\cdot\epsilon_B^*\,\epsilon_i\cdot(p_f+p_i)\right.\bigg]\,,
\end{eqnarray}
and finally, the photon pole contribution
\begin{eqnarray}\!\!\!\!\!\!\!\!\!\!\!\!\!\!\!\!\!\!\!\!\!\!\!\!\!\!\!\!\!\!\!\!\!\!\!\!\!\!\!\!\!\!\!\!\!\!\!\!\!\!\!\!\!\!\!\!\!\!\!\!\!\!\!\!\!\!\!\!\!\!\!\!\!\!\!\!\!\!\!\!\!\!\!\!\!\!\!\!\!
{\rm \gamma-pole-d}:&&\!\!\!\! {{\rm Amp_d}(S=1)}=-{e\,\kappa\over 2k_f\cdot k_i}\nonumber\\
&\times&\left.\bigg[\epsilon_B^*\cdot\epsilon_A\left[\epsilon_f^*\cdot(p_f+p_i)(k_f\cdot
k_i\epsilon_f^*\cdot\epsilon_i-\epsilon_f^*\cdot k_i\,\epsilon_i\cdot k_f)\right.\right.\nonumber\\
&+&\left.\left.\epsilon_f^*\cdot
k_i(\epsilon_f^*\cdot\epsilon_i\, k_i\cdot(p_i+p_f)-\epsilon_f^*\cdot
k_i\epsilon_i\cdot(p_f+p_i))\right]\right.\nonumber\\
&-&\left.2\epsilon_B^*\cdot p_i\left[\epsilon_f^*\cdot\epsilon_A\,(k_f\cdot
k_i\,\epsilon_f^*\cdot\epsilon_i-\epsilon_f^*\cdot
k_i\,\epsilon_i\cdot k_f)\right.\right.\nonumber\\
&+&\left.\left.\epsilon_f^*\cdot
k_i\,(\epsilon_f^*\cdot\epsilon_i\, \epsilon_A\cdot k_i-\epsilon_f^*\cdot
k_i\,\epsilon_i\cdot\epsilon_A)\right]\right.\nonumber\\
&-&\left.2\epsilon_A\cdot p_f\left[\epsilon_f^*\cdot\epsilon_B^*(k_f\cdot
k_i\,\epsilon_f^*\cdot\epsilon_i-\epsilon_f^*\cdot
k_i\,\epsilon_i\cdot k_f)\right.\right.\nonumber\\
&+&\left.\left.\epsilon_f^*\cdot
k_i\,(\epsilon_f^*\cdot\epsilon_i \,\epsilon_B^*\cdot k_i-\epsilon_f^*\cdot
k_i\,\epsilon_i\cdot\epsilon_B^*)\right]\right.\bigg]\,.
\end{eqnarray}

\noindent In the case of gravitational Compton scattering---Figure~\ref{fig:gravcomp}---we have the four contributions\\ \\
\begin{center}
{\bf Gravitational Compton Scattering: spin-1}
\end{center}

\begin{eqnarray}
{\rm Born-a}:&& {\rm Amp}_a(S=1)=\kappa^2\,{1\over 2p_i\cdot
k_i}\bigg[(\epsilon_i\cdot p_i)^2 (\epsilon_f^*\cdot
p_f)^2\epsilon_A\cdot\epsilon_B^*\nonumber\\
&-&(\epsilon_f^*\cdot p_f)^2\epsilon_i\cdot p_i (\epsilon_A\cdot
k_i\epsilon_B^*\cdot \epsilon_i+
\epsilon_A\cdot\epsilon_i\,\epsilon_B^*\cdot p_i)\nonumber\\
&-&(\epsilon_i\cdot p_i)^2\epsilon_f^*\cdot p_f\,
(\epsilon_B^*\cdot\epsilon_f^*\,\epsilon_A\cdot
p_f+\epsilon_B^*\cdot
k_f\,\epsilon_A\cdot\epsilon_f^*)\nonumber\\
&+&\epsilon_i\cdot p_i\,\epsilon_f^*\cdot p_f\,\epsilon_i\cdot
p_f\,\epsilon_A\cdot
k_i\,\epsilon_B^*\cdot\epsilon_f^*+\epsilon_i\cdot
p_i\,\epsilon_f^*\cdot p_f\,\epsilon_f^*\cdot p_i\,\epsilon_A\cdot
\epsilon_i\,\epsilon_B^*\cdot k_f\nonumber\\
&+&(\epsilon_f^*\cdot
p_f)^2\epsilon_B^*\cdot\epsilon_i\,\epsilon_A\cdot\epsilon_i\,
p_i\cdot k_i+(\epsilon_i\cdot
p_i)^2\epsilon_B^*\cdot\epsilon_f^*\,\epsilon_A\cdot\epsilon_f^*\,
p_f\cdot k_f\nonumber\\
&+&\epsilon_i\cdot p_i\,\epsilon_f^*\cdot p_f\,(\epsilon_A\cdot
k_i\,\epsilon_B^*\cdot k_f\,\epsilon_i\cdot
\epsilon_f^*+\epsilon_B^*\cdot\epsilon_f^*\,\epsilon_A\cdot\epsilon_i\,p_i\cdot
p_f)\nonumber\\
&-&\epsilon_i\cdot p_i\,\epsilon_f^*\cdot p_i\,\epsilon_B^*\cdot
\epsilon_f^*\,\epsilon_A\cdot\epsilon_i\,p_f\cdot
k_f-\epsilon_f^*\cdot p_f\,\epsilon_i\cdot
p_f\,\epsilon_A\cdot\epsilon_i\,\epsilon_B^*\cdot\epsilon_f^*\,p_i\cdot
k_i\nonumber\\
&-&\epsilon_i\cdot p_i\,\epsilon_A\cdot k_i\,\epsilon_B^*\cdot
\epsilon_f^*\,\epsilon_f^*\cdot\epsilon_i\,p_f\cdot
k_f-\epsilon_f^*\cdot p_f\,\epsilon_B^*\cdot
k_f\,\epsilon_A\cdot\epsilon_i\,\epsilon_i\cdot\epsilon_f^*\,p_i\cdot
k_i\nonumber\\
&+&\epsilon_A\cdot\epsilon_i\,\epsilon_B^*\cdot\epsilon_f^*\,p_i\cdot
k_i\,p_f\cdot
k_f\,\epsilon_i\cdot\epsilon_f^*-m^2\epsilon_B^*\cdot\epsilon_f^*\,
\epsilon_A\cdot\epsilon_i\,\epsilon_f^*\cdot
p_f\,\epsilon_i\cdot p_i\bigg]\,.\nonumber\\
\end{eqnarray}

\begin{eqnarray}\!\!\!\!\!\!\!\!\!\!\!\!\!\!
{\rm Born-b}:&& {\rm Amp}_b(S=1)=-\kappa^2{1\over 2p_i\cdot
k_f}\bigg[\big(\epsilon_f^*\cdot p_i\big)^2 \big(\epsilon_i\cdot
p_f\big)^2\epsilon_A\cdot\epsilon_B^*\nonumber\\
&+&\big(\epsilon_i\cdot p_f\big)^2\epsilon_f^*\cdot p_i
\big(\epsilon_A\cdot k_f\,\epsilon_B^*\cdot\epsilon_f^*
-\epsilon_A\cdot\epsilon_f^*\,\epsilon_B^*\cdot p_i\big)\nonumber\\
&+&\big(\epsilon_f^*\cdot p_i\big)^2\epsilon_i\cdot p_f \big(\epsilon_B^*\cdot
k_i\epsilon_A\cdot\epsilon_i-\epsilon_B^*\cdot\epsilon_i\,\epsilon_A\cdot p_f\big)\nonumber\\
&-&\epsilon_f^*\cdot p_i\,\epsilon_i\cdot p_f\,\epsilon_f^*\cdot
p_f\,\epsilon_A\cdot
k_f\,\epsilon_B^*\cdot\epsilon_i-\epsilon_f^*\cdot
p_i\,\epsilon_i\cdot p_f\,\epsilon_i\cdot p_i\,\epsilon_A\cdot
\epsilon_f^*\,\epsilon_B^*\cdot k_i\nonumber\\
&-&\big(\epsilon_i\cdot
p_f\big)^2\epsilon_B^*\cdot\epsilon_f^*\,\epsilon_A\cdot\epsilon_f^*\,
p_i\cdot k_f-\big(\epsilon_f^*\cdot
p_i\big)^2\epsilon_B^*\cdot\epsilon_i\,\epsilon_A\cdot\epsilon_i\,
p_f\cdot k_i\nonumber\\
&+&\epsilon_f^*\cdot p_i\,\epsilon_i\cdot p_f\big(\epsilon_A\cdot
k_f\,\epsilon_B^*\cdot k_i\,\epsilon_i\cdot
\epsilon_f^*+\epsilon_B^*\cdot\epsilon_i\,\epsilon_A\cdot\epsilon_f^*\,p_i\cdot
p_f\big)\nonumber\\
&+&\epsilon_f^*\cdot p_i\,\epsilon_i\cdot p_i\,\epsilon_B^*\cdot
\epsilon_i\,\epsilon_A\cdot\epsilon_f^*\,p_f\cdot k_i+\epsilon_i\cdot
p_f\,\epsilon_f^*\cdot
p_f\,\epsilon_A\cdot\epsilon_f^*\,\epsilon_B^*\cdot\epsilon_i\,p_i\cdot
k_f\nonumber\\
&-&\epsilon_f^*\cdot p_i\,\epsilon_A\cdot k_f\,\epsilon_B^*\cdot
\epsilon_i\,\epsilon_i\cdot\epsilon_f^*\,p_f\cdot k_i-\epsilon_i\cdot
p_f\,\epsilon_B^*\cdot
k_i\,\epsilon_A\cdot\epsilon_f^*\,\epsilon_f^*\cdot\epsilon_i\,p_i\cdot
k_f\nonumber\\
&+&\epsilon_A\cdot\epsilon_f^*\,\epsilon_B^*\cdot\epsilon_i\,p_i\cdot
k_f\,p_f\cdot
k_i\,\epsilon_i\cdot\epsilon_f^*-m^2\epsilon_B^*\cdot\epsilon_i\,
\epsilon_A\cdot\epsilon_f^*\,\epsilon_i\cdot
p_f\,\epsilon_f^*\cdot p_i\bigg]\,.\nonumber\\
\end{eqnarray}
\begin{eqnarray}
{\rm Seagull-c}:&&{\rm Amp}_c(S=1)=-{\kappa^2\over 4}\,
\bigg[\big(\epsilon_i\cdot\epsilon_f^*\big)^2\big(m^2-p_i\cdot
p_f\big)\epsilon_A\cdot\epsilon_B^*\nonumber\\
&+&\epsilon_A\cdot
p_f\,\epsilon_B^*\cdot
p_i\,\big(\epsilon_i\cdot\epsilon_f^*\big)^2
+\epsilon_i\cdot p_i\,\epsilon_f^*\cdot
p_f\,\big(2\epsilon_i\cdot\epsilon_f^*\,\epsilon_A\cdot\epsilon_B^*-
2\epsilon_A\cdot\epsilon_2\,\epsilon_B^*\cdot\epsilon_1\big)\nonumber\\
&+&\epsilon_i\cdot p_f\,\epsilon_f^*\cdot
p_i\,\big(2\epsilon_i\cdot\epsilon_f^*\,\epsilon_A\cdot\epsilon_B^*-
2\epsilon_A\cdot\epsilon_i\,\epsilon_B^*\cdot\epsilon_f^*\big)\nonumber\\
&+&2\epsilon_i\cdot p_i\,\epsilon_1\cdot
p_f\,\epsilon_A\cdot\epsilon_f^*\,\epsilon_B^*\cdot\epsilon_f^*+2\epsilon_f^*\cdot
p_f\,\epsilon_f^*\cdot
p_i\,\epsilon_A\cdot\epsilon_i\,\epsilon_B^*\cdot\epsilon_i\nonumber\\
&-&2\epsilon_i\cdot p_i\,\epsilon_i\cdot\epsilon_f^*\,\epsilon_A\cdot
p_f\,\epsilon_B^*\cdot\epsilon_f^*-2\epsilon_f^*\cdot
p_f\,\epsilon_i\cdot\epsilon_f^*\,\epsilon_A\cdot\epsilon_i\,\epsilon_f^*\cdot
p_i\nonumber\\
&-&2\epsilon_i\cdot
p_f\,\epsilon_i\cdot\epsilon_f^*\,\epsilon_A\cdot\epsilon_f^*\,\epsilon_B^*\cdot
p_i-2\epsilon_f^*\cdot
p_i\,\epsilon_i\cdot\epsilon_f^*\,\epsilon_B^*\cdot\epsilon_i\,\epsilon_A\cdot
p_f\nonumber\\
&-&2\big(m^2-p_f\cdot
p_i\big)\epsilon_i\cdot\epsilon_f^*\big(\epsilon_A\cdot\epsilon_i\,\epsilon_B^*\cdot\epsilon_f^*
+\epsilon_A\cdot\epsilon_f^*\,\epsilon_B^*\cdot\epsilon_i\big)\bigg]\,,
\end{eqnarray}
and finally the (lengthy) graviton pole contribution is
\begin{eqnarray}\!\!\!\!\!\!\!\!\!\!\!\!\!\!\!\!\!
 {\rm
\ \ \ \ \  g-pole-d}:&&\!\!\!\!\!\!\!\!\!\!{\rm Amp}_d(S=1)=-{\kappa^2\over 16\,k_i\cdot
k_f}\bigg[\epsilon_B^*\cdot\epsilon_A\Big[\big(\epsilon_i\cdot\epsilon_f^*\big)^2\Big[4k_i\cdot
p_i\,p_f\cdot
k_i+4k_f\cdot p_i\,k_f\cdot p_f\nonumber\\ \nonumber
&-&2\big(p_i\cdot k_i\,p_f\cdot k_f+p_f\cdot k_i\,p_i\cdot k_f\big)+6p_i\cdot
p_f\,k_i\cdot k_f\Big]+4\Big[\big(\epsilon_i\cdot k_f\big)^2\epsilon_f^*\cdot p_f\,\epsilon_f^*\cdot
p_i\\ \nonumber &+&\big(\epsilon_f^*\cdot k_i\big)^2\epsilon_i\cdot p_i\,\epsilon_i\cdot
p_f
+\epsilon_i\cdot k_f\,\epsilon_f^*\cdot k_i\big(\epsilon_i\cdot
p_i\,\epsilon_f^*\cdot p_f+\epsilon_i\cdot p_f\,\epsilon_f^*\cdot
p_i\big)\Big]\nonumber\\
&-&4\epsilon_i\cdot\epsilon_f^*\Big[\epsilon_i\cdot
k_f\big(\epsilon_f^*\cdot
p_i\, p_f\cdot k_f+\epsilon_f^*\cdot p_f\,k_f\cdot p_i\big)
+\epsilon_f^*\cdot k_i\big(\epsilon_i\cdot p_i\,p_f\cdot
k_i+\epsilon_i\cdot p_f\,p_i\cdot k_i\big)\Big]\nonumber\\
&-&4k_i\cdot k_f\,\epsilon_i\cdot\epsilon_f^*\big(\epsilon_i\cdot
p_i\,\epsilon_f^*\cdot p_f+\epsilon_i\cdot p_f\,\epsilon_f^*\cdot
p_i\big)-4p_i\cdot p_f\,\epsilon_i\cdot\epsilon_f^*\,\epsilon_i\cdot
k_f\,\epsilon_f^*\cdot k_i\Big]\nonumber\\
&-&\big(p_i\cdot p_f\,\epsilon_B^*\cdot\epsilon_A-\epsilon_B^*\cdot
p_i\,\epsilon_A\cdot p_f\big)\Big[10\big(\epsilon_i\cdot\epsilon_f^*\big)^2k_i\cdot
k_f+4\epsilon_i\cdot\epsilon_f^*\,\epsilon_i\cdot
k_f\,\epsilon_f^*\cdot
k_i\nonumber\\
&-&4\big(\epsilon_i\cdot\epsilon_f^*\big)^2k_i\cdot
k_f-8\epsilon_i\cdot\epsilon_f^*\,\epsilon_i\cdot
k_f\,\epsilon_f^*\cdot
k_i\Big]+\big(p_i\cdot
p_f-m^2\big)\Big[\big(\epsilon_i\cdot\epsilon_f^*\big)^2\big(4\epsilon_A\cdot
k_i\,\epsilon_B^*\cdot k_i\nonumber \\ \nonumber&+&4\epsilon_A\cdot k_f\,\epsilon_B^*\cdot
k_f-2\big(\epsilon_A\cdot k_i\,\epsilon_B^*\cdot k_f+\epsilon_A\cdot
k_f\,\epsilon_B^*\cdot k_i\big)+6\epsilon_B^*\cdot\epsilon_A\,k_i\cdot
k_f\big)\nonumber\\
&+&4\Big[\big(\epsilon_i\cdot
k_f\big)^2\epsilon_A\cdot\epsilon_f^*\,\epsilon_B^*\cdot\epsilon_f^*+\big(\epsilon_f^*\cdot
k_i\big)^2\epsilon_A\cdot\epsilon_i\,\epsilon_B^*\cdot\epsilon_i
+\epsilon_i\cdot k_f\,\epsilon_f^*\cdot
k_f\big(\epsilon_A\cdot\epsilon_i\,\epsilon_B^*\cdot\epsilon_f^*\nonumber \\ \nonumber
&+&\epsilon_A\cdot\epsilon_f^*\,\epsilon_B^*\cdot\epsilon_i\big)\Big]-4\epsilon_i\cdot\epsilon_f^*\Big[\epsilon_i\cdot
k_f\big(\epsilon_A\cdot\epsilon_f^*\,\epsilon_B^*\cdot
k_f+\epsilon_B^*\cdot\epsilon_f^*\,\epsilon_A\cdot k_f\big)\nonumber\\
&+&\epsilon_f^*\cdot
k_i\big(\epsilon_A\cdot\epsilon_i\,\epsilon_B^*\cdot
k_i+\epsilon_B^*\cdot\epsilon_i\,\epsilon_A\cdot k_i\big)+k_i\cdot
k_f\big(\epsilon_A\cdot\epsilon_i\,\epsilon_B^*\cdot\epsilon_f^*+
\epsilon_B^*\cdot\epsilon_i\,\epsilon_A\cdot\epsilon_f^*\big)\nonumber\\ \nonumber&+&\epsilon_A\cdot\epsilon_B^*\,\epsilon_i\cdot
k_f\,\epsilon_f^*\cdot k_i\Big]\Big]
-2\epsilon_A\cdot
p_f\Big[\big(\epsilon_f^*\cdot\epsilon_i\big)^2\Big[2\epsilon_B^*\cdot k_i\,p_i\cdot
k_i+2\epsilon_B^*\cdot k_f\,p_i\cdot k_f\nonumber \\ \nonumber & +&3\epsilon_B^*\cdot
p_i\,k_i\cdot k_f-\big(\epsilon_B^*\cdot k_i\,p_i\cdot k_f+\epsilon_B^*\cdot
k_f\,p_i\cdot k_i\big)\Big]+2\big(\epsilon_i\cdot
k_f\big)^2\epsilon_B^*\cdot\epsilon_f^*\,\epsilon_f^*\cdot
p_i\nonumber\\ \nonumber&+&2\big(\epsilon_f^*\cdot
k_i\big)^2\epsilon_B^*\cdot\epsilon_i\,\epsilon_i\cdot p_i+2\epsilon_i\cdot k_f\epsilon_f^*\cdot
k_i\big(\epsilon_B^*\cdot\epsilon_i\,\epsilon_f^*\cdot
p_i+\epsilon_i\cdot
p_i\,\epsilon_B^*\cdot\epsilon_f^*\big)\nonumber\\
&-&2\epsilon_i\cdot\epsilon_f^*\Big[\epsilon_i\cdot
k_f\big(\epsilon_B^*\cdot\epsilon_f^*\,p_i\cdot k_f+\epsilon_f^*\cdot
p_i\,\epsilon_B^*\cdot k_f\big)+\epsilon_f^*\cdot k_i\big(\epsilon_B^*\cdot\epsilon_i\,p_i\cdot
k_i+\epsilon_B^*\cdot k_i\,\epsilon_i\cdot p_i\big)\Big]\nonumber\\
&-&2k_i\cdot
k_f\,\epsilon_i\cdot\epsilon_f^*\big(\epsilon_B^*\cdot\epsilon_i\,\epsilon_f^*\cdot
p_i+\epsilon_B^*\cdot\epsilon_f^*\,\epsilon_i\cdot p_i\big)
-2\epsilon_B^*\cdot p_i\,\epsilon_i\cdot\epsilon_f^*\,\epsilon_i
\cdot k_f\,\epsilon_f^*\cdot k_i\Big]\nonumber\\
&-&2\epsilon_B^*\cdot
p_i\Big[\big(\epsilon_f^*\cdot\epsilon_i\big)^2\Big[2\epsilon_A\cdot k_i\,p_f\cdot
k_i+2\epsilon_A\cdot k_f\,p_f\cdot k_f+3\epsilon_A\cdot p_f\,k_i\cdot
k_f\nonumber\\
&-&\big(\epsilon_A\cdot k_i\,p_f\cdot k_f+\epsilon_A\cdot k_f\,p_f\cdot
k_f)\Big]+2\big(\epsilon_i\cdot
k_f\big)^2\epsilon_A\cdot\epsilon_f^*\,\epsilon_f^*\cdot
p_f+2\big(\epsilon_f^*\cdot
k_i\big)^2\epsilon_A\cdot\epsilon_i\,\epsilon_i\cdot p_f\nonumber\\
&+&2\epsilon_i\cdot k_f\,\epsilon_f^*\cdot
k_i\big(\epsilon_A\cdot\epsilon_i\,\epsilon_f^*\cdot p_f+\epsilon_i\cdot
p_f\,\epsilon_A\cdot\epsilon_f^*-2\epsilon_i\cdot\epsilon_f^*\Big[\epsilon_i\cdot
k_f\big(\epsilon_A\cdot\epsilon_f^*\,p_f\cdot k_f\nonumber\\ \nonumber&+&\epsilon_f^*\cdot
p_f\,\epsilon_A\cdot k_f\big)
+\epsilon_f^*\cdot k_i\big(\epsilon_A\cdot\epsilon_i\,p_f\cdot
k_i\nonumber +\epsilon_A\cdot k_i\,\epsilon_i\cdot p_f\big)\Big]\nonumber\\
&-&2k_i\cdot
k_f\epsilon_i\cdot\epsilon_f^*\big(\epsilon_A\cdot\epsilon_i\,\epsilon_f^*\cdot
p_f+\epsilon_A\cdot\epsilon_f^*\,\epsilon_i\cdot p_f\big)
-2\epsilon_A\cdot p_f\,\epsilon_i\cdot\epsilon_f^*\,\epsilon_i\cdot
k_f\,\epsilon_f^*\cdot k_i\Big]\bigg]\,.\nonumber\\
\quad
\end{eqnarray}

\end{document}